\documentclass[superscriptaddress,
reprint,
showpacs,preprintnumbers,
nofootinbib,
amsmath,amssymb,
aps,
prc,
]{revtex4-1}

\usepackage{graphicx}\usepackage{dcolumn}\usepackage{bm}\usepackage{comment}\usepackage{multirow}\usepackage{subfigure}
\usepackage{soul}

\usepackage[usenames,dvipsnames]{color}
\usepackage{hyperref}\usepackage[mathlines]{lineno}\DeclareGraphicsExtensions{.pdf,.png,.jpg}
 \usepackage{lipsum}
\begin{document}
\newcommand{\Q}{CUORE}
\newcommand{\q}{\mbox{CUORE-0}}
\newcommand{\qino}{Cuoricino}
\newcommand{\BB}{\ensuremath{\beta\beta}}
\newcommand{\BBless}{\ensuremath{0\nu\beta\beta}}
\newcommand{\TwoNuBB}{\ensuremath{2\nu\beta\beta}}
\newcommand{\TeO}{\ensuremath{\rm TeO_{2}}}
\newcommand{\ohm}{\ensuremath{\Omega}}
\newcommand{\red}[1]{{\color{red}{#1}}}
\newcommand{\MM}[1]{{\color{blue}{#1}}}
\newcommand{\MMcom}[1]{{\color{OliveGreen}{#1}}}
\newcommand{\black}[1]{{\color{black}{#1}}}
\newcommand{\cky}{counts/(kg$\cdot$yr)}
\newcommand{\ckky}{counts/(keV$\cdot$kg$\cdot$yr)}
\newcommand{\kgyr}{kg$\cdot$yr}
\newcommand{\mbb}{\ensuremath{m_{\beta\beta}}}
\newcommand{\QINOPaper}{CINO2011}
\newcommand{\JOcom}[1]{{\color{red}{#1}}}
\newcommand{\vet}[1]{{\color{magenta}{#1}}}

\newcommand{\edit}[2][]{{#2}}

\newcommand{\BoDs}{BoDs}
\newcommand{\bd}{i}

\newif\ifisgrayselect

\isgrayselectfalse

\newcommand{\grayselect}[2]{\ifisgrayselect#1\else#2\fi}

 \newcommand{\PaperRevision}{973:1025M}

\grayselect{\graphicspath{{images/}{images/bw/}}}{\graphicspath{{images/}{images/color/}}}

\title{Analysis Techniques for the Evaluation of the Neutrinoless Double-$\beta$ Decay Lifetime in $^{130}$Te with CUORE-0}
\newcommand{\deceased}{\altaffiliation {Deceased.}}
\newcommand{\atprincetonnow}{\altaffiliation {Present address: Physics Department, Princeton University, Princeton, NJ 08544, USA}}
\newcommand{\correspondence}{\email {e-mail: cuore-spokesperson@lngs.infn.it}}

\author{C.~Alduino}
\affiliation{ Department of Physics and Astronomy, University of South Carolina, Columbia, SC 29208 - USA }

\author{K.~Alfonso}
\affiliation{ Department of Physics and Astronomy, University of California, Los Angeles, CA 90095 - USA }

\author{D.~R.~Artusa}
\affiliation{ Department of Physics and Astronomy, University of South Carolina, Columbia, SC 29208 - USA }
\affiliation{ INFN - Laboratori Nazionali del Gran Sasso, Assergi (L'Aquila) I-67010 - Italy }

\author{F.~T.~Avignone~III}
\affiliation{ Department of Physics and Astronomy, University of South Carolina, Columbia, SC 29208 - USA }

\author{O.~Azzolini}
\affiliation{ INFN - Laboratori Nazionali di Legnaro, Legnaro (Padova) I-35020 - Italy }

\author{T.~I.~Banks}
\affiliation{ Department of Physics, University of California, Berkeley, CA 94720 - USA }
\affiliation{ Nuclear Science Division, Lawrence Berkeley National Laboratory, Berkeley, CA 94720 - USA }

\author{G.~Bari}
\affiliation{ INFN - Sezione di Bologna, Bologna I-40127 - Italy }

\author{J.W.~Beeman}
\affiliation{ Materials Science Division, Lawrence Berkeley National Laboratory, Berkeley, CA 94720 - USA }

\author{F.~Bellini}
\affiliation{ Dipartimento di Fisica, Sapienza Universit\`{a} di Roma, Roma I-00185 - Italy }
\affiliation{ INFN - Sezione di Roma, Roma I-00185 - Italy }

\author{A.~Bersani}
\affiliation{ INFN - Sezione di Genova, Genova I-16146 - Italy }

% ...
 \collaboration{CUORE Collaboration}\date{April 25, 2016}                            \begin{abstract}
We describe in detail the methods used to obtain the lower bound on
the lifetime of neutrinoless double-beta (\BBless) decay in $^{130}$Te
and the associated limit on the effective Majorana mass of the
neutrino using the {\q} detector. {\q} is a bolometric detector array
located at the Laboratori Nazionali del Gran Sasso that was designed
to validate the background reduction techniques developed for {\Q}, a
next-generation experiment scheduled to come online in 2016. {\q} is
also a competitive {\BBless} decay search in its own right and
functions as a platform to further develop the analysis tools and
procedures to be used in {\Q}. These include data collection, event
selection and processing, as well as an evaluation of signal
efficiency.  In particular, we describe the amplitude evaluation,
thermal gain stabilization, energy calibration methods, and the
analysis event selection used to create our final {\BBless} search
spectrum. We define our high level analysis procedures, with emphasis
on the new insights gained and challenges encountered. We outline in
detail our fitting methods near the hypothesized {\BBless} decay peak
and catalog the main sources of systematic uncertainty. Finally, we
derive the {\BBless} decay half-life limits previously reported for
{\q}, \mbox{$T^{0\nu}_{1/2}>2.7\times10^{24}\,\rm{yr}$}, and in
combination with the {\qino}
limit, \mbox{$T^{0\nu}_{1/2}>4.0\times10^{24}\,\rm{yr}$}.
 \end{abstract}

\maketitle

\section{Introduction}
Neutrinoless double-beta (\BBless) decay~\cite{Furry1939} is a
hypothesized second-order weak decay in which a nucleus simultaneously
converts two neutrons into two protons and produces only two electrons
in the process, \mbox{$(Z,A)\rightarrow(Z+2,A)+2\beta^-$}. The
discovery of this decay would conclusively indicate that lepton number
is violated and that neutrinos are Majorana fermions. Further, it
could help constrain the absolute scale of the neutrino masses and
their hierarchy \cite{Avignon2008}, and would lend support to the
theory that neutrinos played a fundamental role in the creation of the
matter asymmetry of the
Universe~\cite{Feinberg1959,Pontecorvo1968}. For all these reasons,
the search for {\BBless} decay has recently become the center of
intense experimental effort utilizing a broad range of technologies
\cite{Giuliani2012,Cremonesi2013,Barabash2015}.  At present, {\BBless}
decay has never been decisively observed, but several recent
experiments have placed upper limits on its decay rate in $^{76}$Ge
\cite{GERDA2013}, $^{136}$Xe~\cite{EXO2014,KamLANDZen2013} and
$^{130}$Te~\cite{Q0FinalPrl}.

The Cryogenic Underground Observatory for Rare Events (\Q)
\cite{CUORE_NIMA,CUORE_arxiv} is a next-generation tonne-scale
bolometric detector, currently in the advanced stages of construction
at the Laboratori Nazionali del Gran Sasso (LNGS) of INFN and expected
to begin operation in 2016. {\Q} will search for the {\BBless} decay
of $^{130}$Te using a segmented array of 988 TeO$_2$ bolometric
detectors operated at extremely low temperatures. The detectors will
be arranged into an array of 19 towers with 52 bolometers each for a
total detector mass of 741\,kg, or 206\,kg of $^{130}$Te.

{\Q} builds on the experience of {\qino}~\cite{Cuoricino_PLB,
  Cuoricino_PRC, QINOPaper}, which was a single tower of 62 bolometers
with a total mass of $\sim$40\,kg. {\qino} ran from 2003--2008 and
until recently held the best limits on the {\BBless} decay half-life
of $^{130}$Te at $T^{0\nu}_{1/2}>2.8\times10^{24}\,\rm{yr}$ (90\% C.L.)
\cite{QINOPaper}. Moving from
{\qino} to {\Q}, we seek to increase the active mass and improve
sensitivity to {\BBless} decay by reducing backgrounds through better
material cleaning and handling~\cite{TTTPaper, CuoreXtals, CCVR}. The
{\q} detector is a single {\Q}-style tower, with comparable active
mass to {\qino}, that was operated in the {\qino} cryostat from
2013 to 2015.  {\q} serves as a technical prototype and validation of
the background reduction techniques developed for \Q, as well as a
sensitive {\BBless} decay search on its own.

This paper begins by briefly describing the design and operation of
the {\q} detector in Section~\ref{sec:DetectorDesign}; a more detailed
and technical description of the detector design and performance can
be found in~\cite{Q0DetectorPaper}, and a report on the initial
performance can be found in~\cite{Q0InitialPaper}. In
Section~\ref{sec:DataProduction}, we describe the production of the
{\q} energy spectrum; this process closely follows the one used for
{\qino} found in~\cite{QINOPaper} (hereafter referred to as
\QINOPaper), so here we focus on the parts of the analysis that have
been further developed for {\q}, including the new data blinding
procedure. In Section~\ref{sec:CutsAndEfficiency}, we outline the data
selection criteria and the signal efficiency
evaluation. Section~\ref{sec:DataUnblinding} summarizes our unblinding
procedure. In Section~\ref{sec:NDBDAnalysis}, we present our technique
for searching for a {\BBless} decay signal and derive the limit on the
half-life of {\BBless} decay of $^{130}$Te previously presented in
\cite{Q0FinalPrl}. In Section~\ref{sec:DetectorPerformance}, we detail
the performance of the {\q} detector, particularly in comparison to
the {\qino} detector. In Section~\ref{sec:CombinedLimits} we present
the technique for combining the results of {\q} and {\qino} to obtain
the limit on the {\BBless} decay half-life of $^{130}$Te presented in
\cite{Q0FinalPrl}.
 
\section{Detector Design \& Data Collection}
\label{sec:DetectorDesign}

\begin{figure}[h]
  \begin{minipage}{.155\columnwidth}
    \subfigure[]{\label{fig:tower}\includegraphics[width=\linewidth]{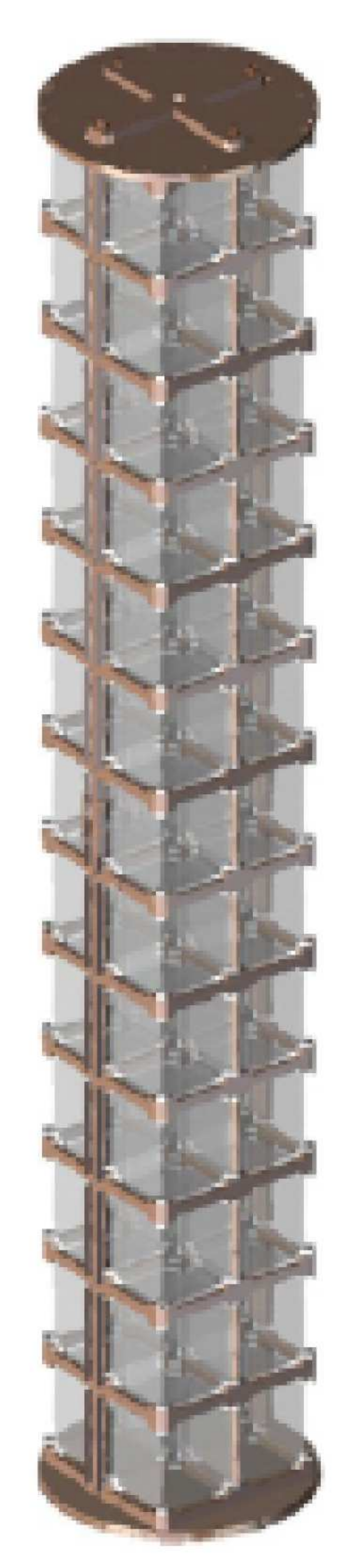}}
  \end{minipage}
  \begin{minipage}{.83\columnwidth}
    \subfigure[]{\label{fig:illustration}      \includegraphics[width=0.51\linewidth]{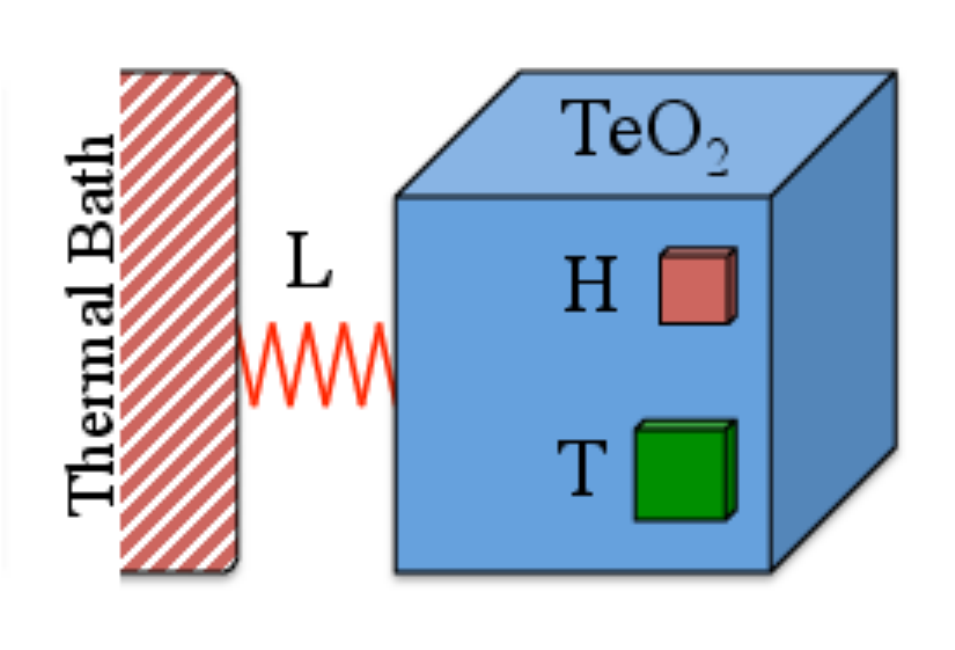}    }
    \subfigure[]{\label{fig:bol_sig}      \includegraphics[width=0.90\linewidth]{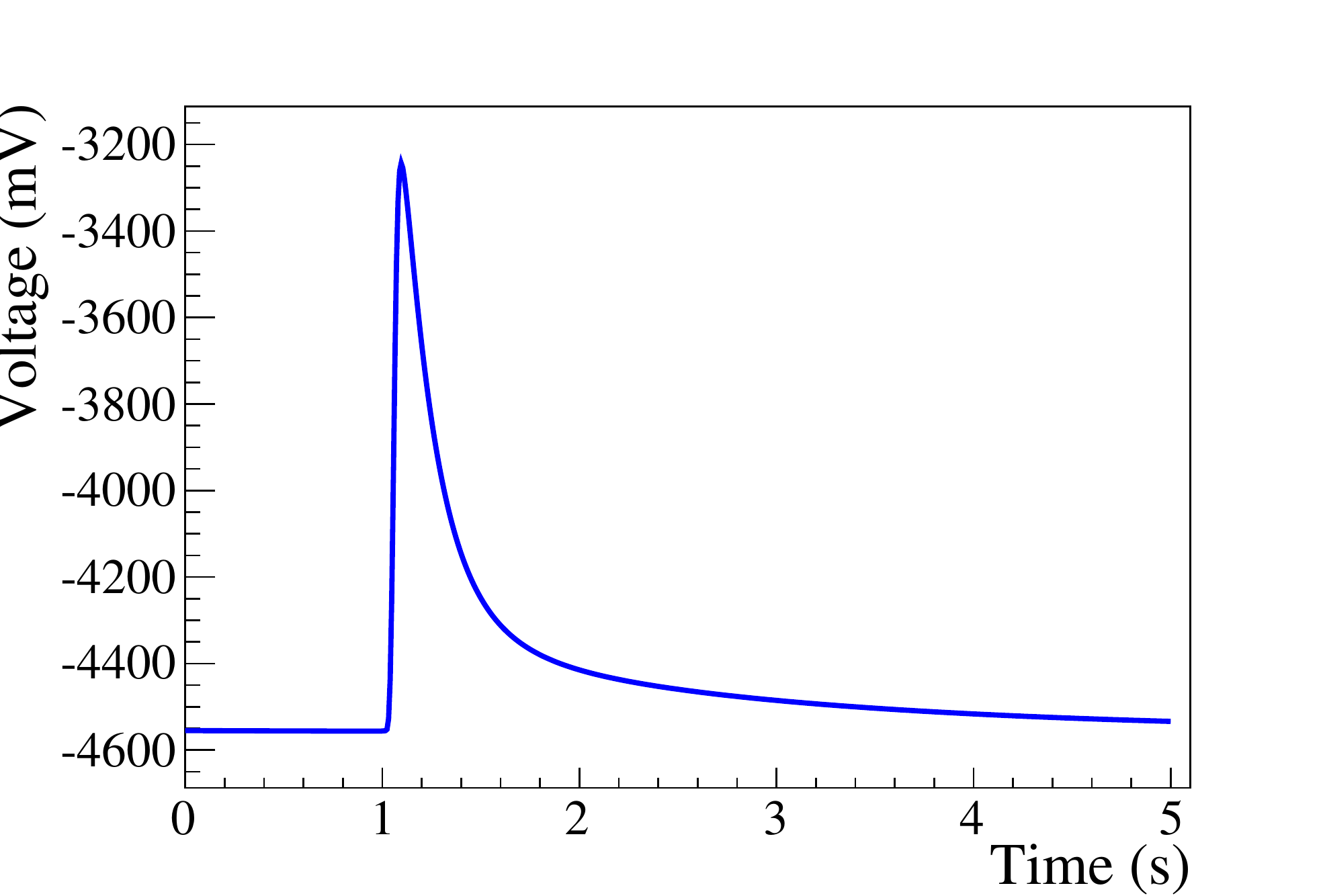}    }
  \end{minipage}
  \caption{(\emph{a}) {\q} tower array rendering. The tower consists
    of 13 floors of 4 bolometers, mounted in a copper
    frame. (\emph{b}) Schematic of a single {\q} bolometer showing the
    thermistor (T), the heater (H), and the weak thermal link (L)
    between TeO$_2$ crystal and copper thermal bath (not to
    scale). (\emph{c}) An example of a bolometer signal with an energy
    of approximately 2615~keV. The rise and fall times of this signal
    are 0.05\,s and 0.2\,s respectively. Figure
    from~\cite{Q0InitialPaper}.}
\end{figure}

The {\q} experiment is a segmented array of 52 bolometric detectors
arranged into a tower of 13 floors with 4 bolometers per floor (see
Fig.~\ref{fig:tower}). Each bolometer has three primary components: an
energy absorber, a temperature sensor, and a weak thermal link to the
copper frame that acts both as the structural tower support and the
thermal bath (see Fig.~\ref{fig:illustration}). When energy is
deposited in the absorber, its temperature increases suddenly by
\begin{equation}
  \delta T = E/C(T),
\end{equation}
where $C(T)$ is the (temperature-dependent) heat capacity and $E$ is
the amount of energy deposited. As the energy slowly leaks out into
the thermal bath, the bolometer gradually returns to its initial
temperature. This temperature pulse is converted to a voltage pulse by
the thermometer (see Fig.~\ref{fig:bol_sig}) and by measuring its
amplitude we can determine the amount of energy deposited in the
bolometer.

In {\q}, the energy absorber is a 750\,g
\mbox{$5\times5\times5\,\rm{cm}^{3}$} $^{\rm nat}$TeO$_2$ crystal
which we cool to an operating temperature of
$T_0\approx12\,\rm{mK}$. The typical heat capacity at this temperature
corresponds to \mbox{$\Delta T/\Delta E\sim10 - 20\,\mu\rm{K/MeV}$}. The
natural isotopic abundance of $^{130}$Te is
$a_I=34.167$\%~\cite{Fehr2004}, thus the crystal acts as both the
source of the decays of interest and detector of their energy. In this
``source = detector'' configuration, Monte Carlo simulations show that
$\approx88\%$ of {\BBless} decay events deposit all of their energy in
the crystal in which the decay occurred. Thus the signal we are
searching for is a monoenergetic peak at the Q-value of the $^{130}$Te
decay,
\mbox{$Q_{\beta\beta}=2527.518\pm0.013\,\rm{keV}$}~\cite{Redshaw2009}. This
energy is above that of the majority of the naturally occurring
environmental $\gamma$ radiation, but between the prominent 2615~keV
line from the decay of $^{208}$Tl and its Compton edge. In this
region, the primary backgrounds are due to multi-scattered $\gamma$
events and degraded $\alpha$ decays which reach our detectors from the
surfaces of materials near the crystals. A detailed description of the
relevant backgrounds can be found in~\cite{Q0BackgroundModel}.

The {\q} tower has a total active mass of 39.1~\,kg for a total
$^{130}$Te mass of 10.9\,kg. The tower is cooled in the cryostat that
housed the {\qino} experiment. The cryogenic installation, shielding,
and anti-radon system are identical to {\qino}
(see~\cite{Q0DetectorPaper} for details) and the backgrounds
associated with this infrastructure is similarly unchanged (see
Sec.~\ref{sec:DetectorPerformance}).

We monitor the temperature of each bolometer by measuring the
resistance of a neutron transmutation doped (NTD) Ge thermistor glued
to each crystal. The NTD has a resistivity that is exponentially
dependent on its temperature, making it a very sensitive
thermometer~\cite{Haller1984,Fiorini1984,Enss2008}. We further
instrument each crystal with a silicon resistor, which we use as a
Joule heater to produce fixed-energy reference pulses for stabilizing
the gain of the bolometers against temperature variations. Each
bolometer is held in the copper frame with a set of
polytetrafluoroethylene (PTFE) supports. These, as well as the
25\,$\mu$m gold wires that instrument the NTD and Joule heater, form
the weak thermal link to the thermal bath.

We bias each NTD through two low-noise load resistors and measure the
output voltage signal using a specially designed low-noise room
temperature preamplifier, a programmable gain amplifier, and a 6-pole
Thomson-Bessel low-pass filter with a programmable cutoff frequency
set to 12\,Hz. The data-acquisition system (DAQ) continuously samples
each waveform at 125\,S/s with $\pm$10.5\,V dynamic range and 18\,bit
resolution. We trigger each data stream in software and store events
in 5 second windows: the one second of data preceding the trigger and
the four seconds after. Particle pulses --- pulses coming from energy
deposits in the crystals --- have typical rise \edit{times of
  $\sim$0.05\,s and two decay time components, a fast decay time of
  $\sim$0.2\,s and a slower decay time of $\sim$1.5\,s. The former
  decay time is determined by the heat capacity of the crystal and the
  thermal conductivity to the thermal bath, and the latter decay time
  by the heat capacity of the of the auxiliary components (i.e. the
  PTFE spacers and nearby copper frame). The rise time is determined
  primarily by the roll-off of the Bessel filter.} Typical trigger
thresholds range from 30\,keV to 120\,keV. Every 200\,s, we collect 5
second waveforms simultaneously on all channels with no signal trigger
and use these to study the noise behavior of the detector.

We collect data in one-day-long runs, which are interrupted for
2-3~hours every 48~hours to refill the liquid He bath and perform
other maintenance on the cryogenic system. Roughly once per month, we
calibrate the energy response of the detector by inserting thoriated
tungsten wires inside the external lead shielding and using the
characteristic $\gamma$ lines from the $^{232}$Th decay chain. These
calibration runs typically last for three days.  The data are combined
into datasets that contain roughly three weeks of {\BBless} decay
physics runs flanked at the beginning and the end by a set of
calibration runs. Each crystal has a typical event rate of $\sim$1\,mHz
in the physics runs and $\sim$60\,mHz in the calibration.

During the tower assembly, one NTD and one heater could not be bonded,
and another heater was lost during the first cool down. Thus of the 52
bolometers, 49 are fully instrumented (working heater and thermistor),
two are functional but without heater (thermistor only), and one
cannot be read (no thermistor).

The detector was assembled in March 2012 and first cooled down in
August 2012, with data collection starting in March 2013. The first
data-taking campaign (Campaign I) lasted until September 2013
(8.5\,{\kgyr} of {\TeO}, corresponding to 2.0\,{\kgyr} of $^{130}$Te).
We then paused data collection for about 2 months to perform
maintenance on the cryostat. Data collection resumed in November 2013
and Campaign II lasted until March 2015. Combining both campaigns, the
total exposure is 35.2\,{\kgyr} of TeO$_2$, corresponding to
9.8\,{\kgyr} of $^{130}$Te.
 
\section{First-Level Data Processing}
\label{sec:DataProduction}

The low-level data processing takes the {\q} data from a series of
triggered waveforms to a calibrated energy spectrum that will be the
input into the higher-level analysis. The data processing procedure
for {\q} closely follows that of {\qino}, outlined in \QINOPaper, but
with several additions newly developed for {\q}.

In order to estimate the energy of each event, we model the
time-waveform $v_i(t)$ of each bolometer, $i$, as the sum of a known
detector response function $s_i(t)$ and an unknown additive noise
term $n_i(t)$
\begin{equation}
  v_i(t)=B_i\,s_i(t)+n_i(t),
\end{equation}
where $B_i$ is the amplitude of the signal response. To a very good
approximation, this amplitude can be decomposed as
\begin{equation}
  B_i=G_i(T)\cdot A_i(E),
\end{equation}
where $A_i(E)$ depends only on the energy deposited into the
bolometer $E$, and $G_i(T)$ is a bolometric gain which depends on the
operating temperature of the bolometer $T$. The low-level data
processing performs the following steps on each triggered waveform in
order to extract the deposited energy:
\begin{enumerate}
\item measure the amplitude of the signal $B_i$ while minimizing the
  effect of the noise term in order to maximize the energy resolution
  of our detector (pulse amplitude evaluation);
  \label{item:amplitude}
\item stabilize the temperature-dependent gain term $G_i(T)$ against
  temperature drifts of the detector (thermal gain stabilization);
  \label{item:stabilization}
\item determine an energy calibration that models the form of $A_i(E)$,
  allowing us to extract the energy for each event (energy
  calibration);
  \label{item:calibration}
\item blind the region of interest (ROI) in order to prevent any bias
  in the later stages of our analysis (data blinding).
  \label{item:blinding}
\end{enumerate}

\begin{figure}
  \centering
  \includegraphics[width=.98\columnwidth]{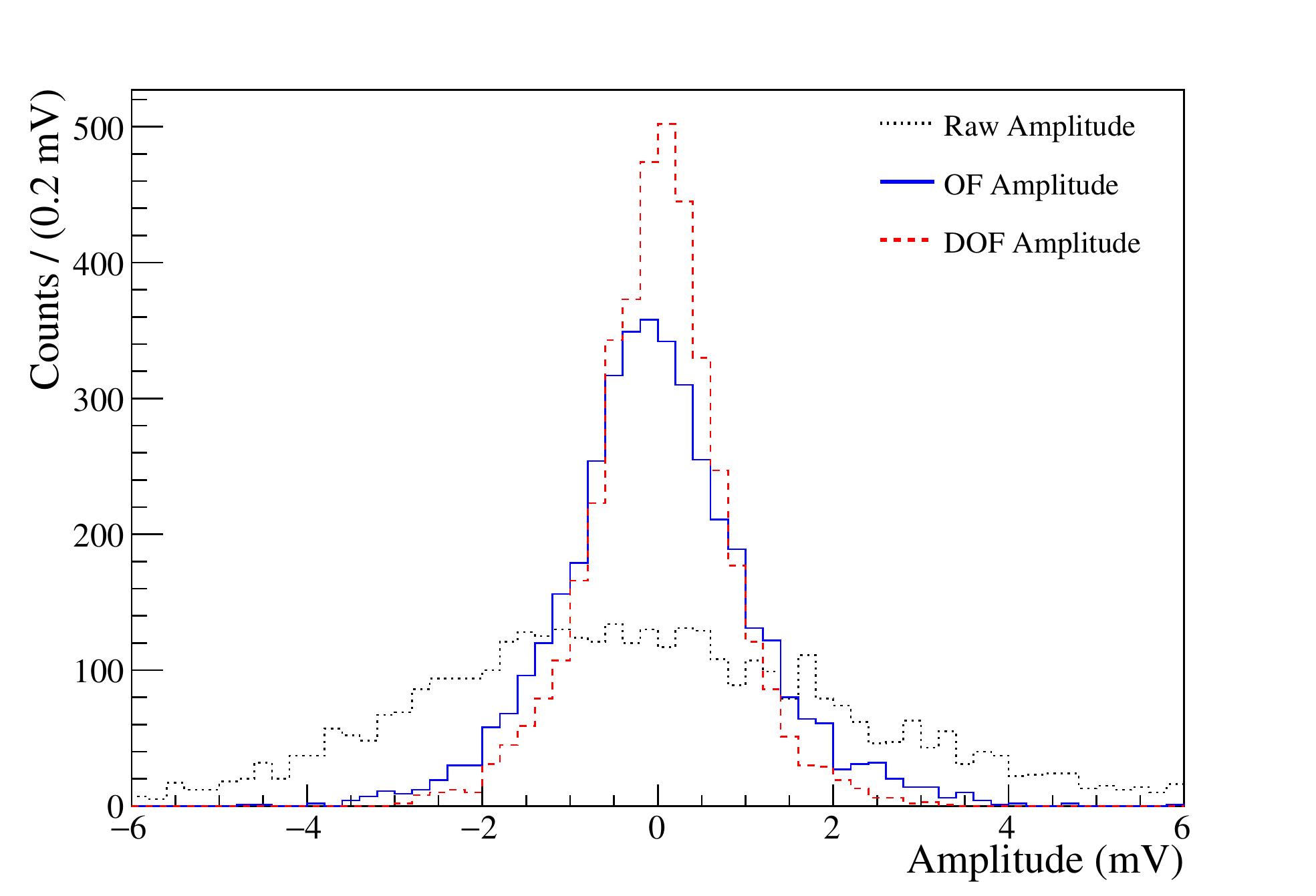}
  \caption{The distribution of amplitudes of the noise pulses
    collected from a single channel during the physics data of a
    dataset from Campaign I. The widths of the above distributions are
    indicative of the amount of noise remaining after filtering. The
    channel presented is one where the DOF performed well. The raw
    unfiltered RMS is 2.7\,mV (\emph{\grayselect{dotted}{black dotted}
      histogram}); the RMS after OF is 1.1\,mV
    (\emph{\grayselect{solid}{blue solid} histogram}); the RMS after
    DOF is 0.8\,mV (\emph{\grayselect{dashed}{red dashed}
      histogram}). }
  \label{fig:FilterNoiseComparison}
\end{figure}

\textbf{\ref{item:amplitude}. Amplitude Evaluation:} To evaluate the
amplitude of the pulse $B_i$, we employ two parallel approaches. We
apply the same optimum filtering (OF) technique described
in~\cite[\QINOPaper,][]{Gatti1986} as well as a new decorrelating
optimum filter (DOF). Both filters are frequency-based and designed to
maximize the signal-to-noise ratio (SNR), assuming a predetermined
detector response function $s_i(t)$ and noise spectrum (see
Fig.~\ref{fig:FilterNoiseComparison}). These filters leverage the
entire waveform to create an amplitude estimate rather than just a few
points around the peak of the pulse.

Up to a multiplicative gain, an OF pulse can be written in frequency
space as
\begin{equation}
  V^{\rm OF}_i(\omega)\propto e^{i\omega t_{\rm
      max}}\frac{S^*_i(\omega)}{N_i(\omega)}V_i(\omega),
  \label{eqn:OFTransferFunction}
\end{equation}
where \edit{$V_i(\omega)$ and} $S_i(\omega)$ is the Fourier transform
of the \edit{signal $v_i(t)$ and} temporal detector response function
$s_i(t)$ respectively for bolometer $i$; $N_i(\omega)$ is the noise
power spectral density of the underlying noise sources; $\omega$ is
the angular frequency and $t_{\rm max}$ is the time at which the pulse
reaches its maximum. The expected detector response $s_i(t)$ is
computed for each bolometer over each dataset by averaging many events
in the 2615\,keV calibration line. The exact number of events depends
on the counting rate and the amount of calibration data in a given
dataset, but is typically several hundred events. The noise power
spectral density $N_i(\omega)$ is similarly estimated for each
bolometer on each dataset by averaging the noise power spectral
densities of noise samples collected throughout each run.

The DOF generalizes Eqn.~(\ref{eqn:OFTransferFunction}) by accounting
for noise correlations between neighboring
bolometers~\cite{Mancini-Terracciano2012,Ouellet2015}. The DOF pulse
for an event on bolometer $i$ is given by
\begin{equation}
  V_i^{\rm DOF}(\omega)\propto e^{i\omega t_{\rm max}}\sum_{j} S^*_i(\omega)C^{-1}_{ij}(\omega)V_j(\omega),
  \label{eqn:DOFTransferFunction}
\end{equation}
where $C^{-1}_{ij}(\omega)$ is the $i,j$ component of the inverted
noise covariance matrix at frequency $\omega$ and the sum runs over a
list of correlated bolometers. In {\q}, we limit this list to the 11
nearest geometric neighbors for bolometers in the middle floors of the
tower (i.e., the four bolometers from the floor above, the four from
the floor below, and the three on the same floor as the triggered
bolometer) or the 7 nearest neighbors for bolometers on the top and
bottom floors. This filter can be thought of as working in two stages:
it first subtracts the noise common to all bolometers and then
performs a regular OF on the bolometer of interest with the expected
noise spectrum after removing common-mode noise. The key is that the
neighboring bolometers provide an estimate of the common-mode
noise. Note that if the covariance matrix is calculated with only the
bolometer of interest and no neighboring bolometers (i.e., if
$C_{ij}(\omega)$ is diagonal), then
Eqn.~(\ref{eqn:DOFTransferFunction}) reduces to
Eqn.~(\ref{eqn:OFTransferFunction}).

The DOF typically outperforms the OF in reducing the RMS of the noise
in the physics runs but performs worse in the calibration runs. The
higher event rate of the calibration runs leads to a higher
probability of an event occurring on a neighboring bolometer within
the 5\,s triggered window which yields an incorrect estimate of the
common-mode noise. This results in two scenarios: either the energy
deposited is small (i.e., not much above the noise), the pulse goes
untriggered, and is inadvertently included in the sum in
Eqn.~(\ref{eqn:DOFTransferFunction}); or the event is triggered and the
waveform is excluded from the sum and the filter is no longer
``optimal'' (i.e., the terms in the sum are not optimized for the
smaller set of bolometers). Both scenarios degrade the performance of
the DOF.

This effect is only prominent in the calibration runs where the event
rate is about 60 times higher than in the physics runs and thus, in
theory, does not worsen the DOF performance on the physics data and in
our {\BBless} analysis. However, as we show in
Section~\ref{sec:NDBDAnalysis}, the calibration runs are essential to
determining the energy resolution input to our {\BBless} decay
analysis, so this makes the DOF problematic. Despite this, for some
bolometers, the benefit of the decorrelation outweighs the degradation
due to the higher event rate in the calibration data. Thus the final
{\q} dataset utilizes both the OF and the DOF, depending on which
performed better on the 2615\,keV $^{208}$Tl line in the calibration
data. In order to use the DOF over the OF the improvement in energy
resolution at 2615\,keV must be statistically significant at the
$\gtrsim$90\% level. With this requirement, 20\% of the final {\q}
data production utilizes the DOF.

Once filtered, the amplitude of each pulse is determined by
interpolating the three data points around the peak of the filtered
pulse and evaluating the maximum of that parabola.

\textbf{\ref{item:stabilization}. Thermal Gain Stabilization:} The
thermal gain stabilization (TGS) compensates slow variation in the
gain of the bolometers $G_i(T)$ due to drifts of the operating
temperature of the detector. As with the amplitude evaluation, we use
two techniques in parallel: a heater-based TGS and a calibration-based
TGS.

The heater-TGS is identical to the technique described in {\QINOPaper}
and described further in~\cite{Alessandrello1998}.  This approach uses
the heater attached to each bolometer to inject fixed-energy reference
pulses every 300\,s during each run. Since the energy of the reference
pulse is constant, any variation in its measured amplitude $B^{\rm
  ref}_i$ is due to a change in the bolometric gain $G_i(T)$. We use
the average value of the baseline, measured in the one second of data
preceding the trigger, as a proxy for the bolometer temperature at the
time of the event. By regressing the reference amplitude $B^{\rm
  ref}_i$ as a function of the baseline value, we can determine
$G_i(T)$ --- up to a multiplicative constant that can be folded into
$A_i(E)$. We then factor $G_i(T)$ out of the measured amplitude $B_i$
to stabilize our bolometric response against thermal drifts.

\begin{figure}
  \centering
  \includegraphics[width=.98\columnwidth]{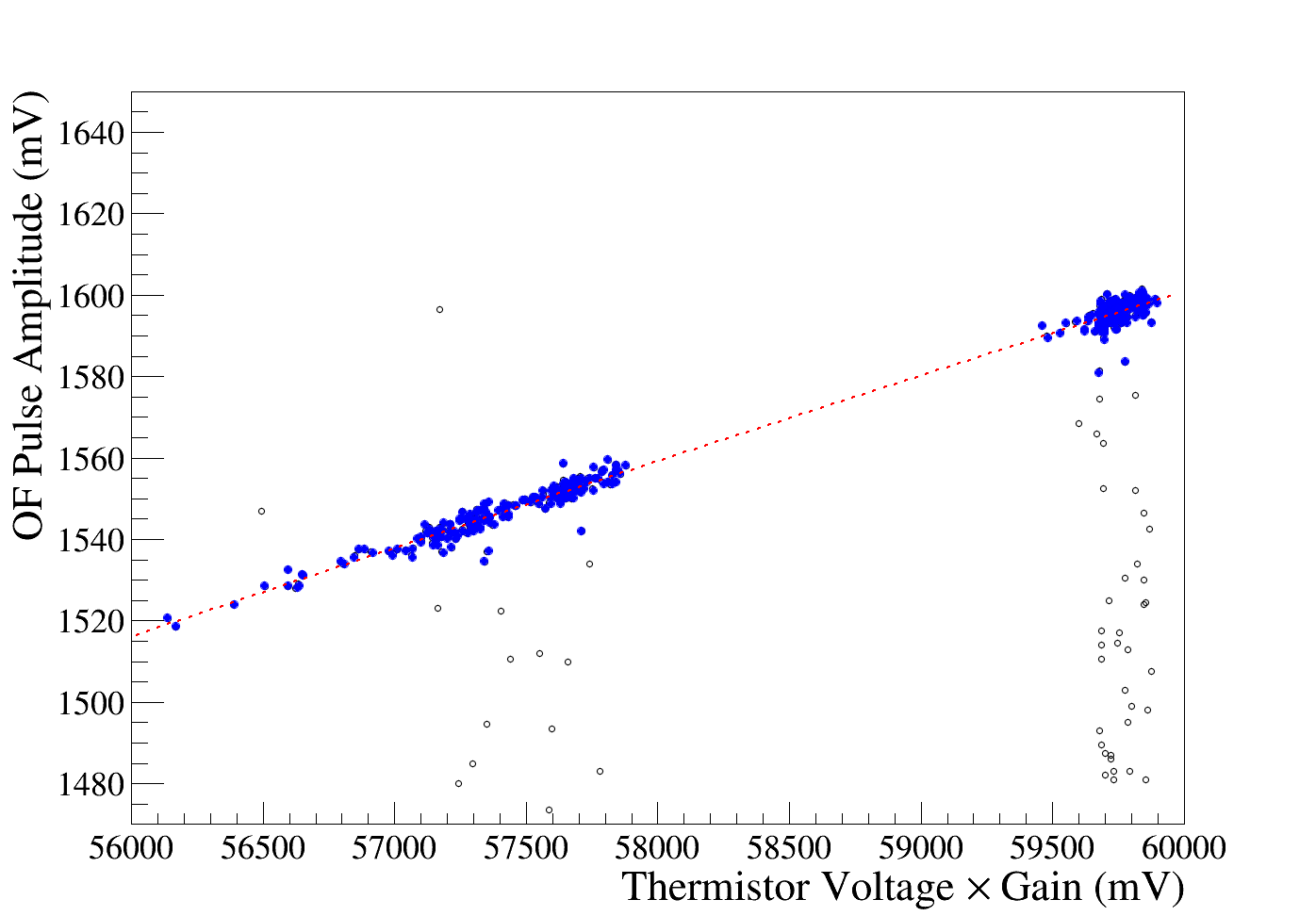}
  \caption{Example of the calibration-TGS. The points are taken during
    the calibration runs for one of the {\q} datasets. The cluster of
    points on the right are from the calibration runs which opened the
    dataset, while the cluster on the left are from the closing
    calibration runs. The solid \grayselect{gray}{blue} points have
    energies around the 2615\,keV $^{208}$Tl peak and are used for
    calibration-TGS. By regressing the measured amplitudes of these
    points against the NTD voltage we can determine a stabilization
    curve (\emph{\grayselect{gray}{red} dashed line}) which is then
    applied to the physics runs taken between calibrations.}
  \label{fig:WoHStabilization}
\end{figure}

For the two bolometers without functioning pulser heaters the
heater-TGS can not be applied. These two bolometers amount to about
4\% of our total exposure. Moreover, for some bolometers the
heater-TGS algorithm consistently failed to stabilize the gain over
very large temperature drifts. This was due partly to deviations from
linearity and partly to differences in the way energy is deposited by
particle interactions versus heater pulses (i.e., differences in the
pulse shapes resulting from particle interactions and heater
pulses). A failure of the heater-TGS manifests as a shift in the
location of the calibration peaks between the initial and final
calibration runs, visible as two distinct peaks in the calibration
spectrum. In this case, we consider the entire dataset invalid for
that particular bolometer. These shifted calibration datasets
correspond to about 7\% of our total exposure.

In order to address these issues, we developed a TGS algorithm based
on calibration data and independent from the heater. This approach
uses the 2615\,keV $\gamma$-line in the calibration runs in lieu of
the monoenergetic pulser to map the temperature-dependent gain
$G_i(T)$. We regress the gain dependence measured in the calibration
runs (see Fig.~\ref{fig:WoHStabilization}) and use this to correct the
amplitudes of events in both the calibration and physics runs. Since
calibration-TGS is interpolated across an entire dataset, it requires
carefully measuring and accounting for the applied and stray voltage
offsets. This calibration-TGS allowed us to recover about 80\% of the
lost exposure on the two bolometers with broken heaters. Additionally,
in cases of large temperature drifts, the calibration-TGS routinely
outperformed the heater-TGS and resulted in little or no shift between
the peak positions in the initial and final calibration runs, as
measured on the 2615\,keV line. This allowed us to recover much of the
7\% of exposure that would have been rejected with the heater-TGS; and
further it improved the resolution of other bolometers that showed a
marginal peak shift between initial and final calibration runs, but
one not large enough to be considered invalid. All told, we used the
calibration-TGS on 12\% of the total {\q} exposure.

For the majority of the {\q} data, applying the TGS caused
temperature-dependent gain drifts to become a subleading cause of
degradation in the energy resolution of our detector. However, in
2.7\% of the final exposure both TGS algorithms failed significantly,
usually due to an abnormally large or sudden drift in temperature or a
change in electronic operating conditions. These data were discarded
for the rest of the analysis.

\textbf{\ref{item:calibration}. Energy Calibration:} For each dataset,
we calibrate the energy response of each bolometer $A_i(E)$ using the
reconstructed positions of at least four of the seven strongest
$\gamma$ peaks from the $^{232}$Th decay chain. This consists of
fitting each peak position using a Gaussian lineshape plus a first
degree polynomial background and performing a linear regression on the
expected energies of the calibration peaks against their reconstructed
positions using a second-order polynomial with zero intercept.

In Section~\ref{sec:NDBDAnalysis}, we show that a Gaussian line shape
does not provide a good fit to the reconstructed peak shapes. This
discrepancy leads to a small bias in the reconstructed event energies,
but rather than correcting for this bias at the calibration stage, we
adjust the position at which we search for a {\BBless} decay signal
(this is detailed in Section~\ref{sec:NDBDAnalysis}). For {\Q}, we
plan to improve our energy reconstruction by accounting for these
non-Gaussian peak shapes during the data processing.

\textbf{\ref{item:blinding}. Data Blinding:} The final step of the
first level data processing is the blinding of the ROI. Our blinding
procedure is designed to mask any possible signal or statistical
fluctuation at $Q_{\beta\beta}$, while maintaining the spectral
integrity so that we can use the blinded energy spectrum for testing
our later analyses. We use a form of data salting that randomly shifts
the reconstructed energy of a fraction of events from within 10\,keV of
the $^{208}$Tl 2615\,keV peak by $-87$\,keV to around $Q_{\beta\beta}$
and the same fraction of events from within 10\,keV of $Q_{\beta\beta}$
by $+87$\,keV to around the $^{208}$Tl peak. Because there are
significantly more events around the $^{208}$Tl peak, this creates an
artificial peak at $Q_{\beta\beta}$ with the shape of a true signal
peak. The fraction of events is blinded and random but chosen from a
range such that the artificial peak is unrealistically large (see
Fig.~\ref{fig:BlindedUnblinded}). Each event's true energy is
encrypted and stored, to be decrypted later during unblinding. This
procedure is pseudo-random and repeatable. The calibration runs are
not blinded. The steps for unblinding are detailed in
Section~\ref{sec:DataUnblinding}.

\section{Data Selection \& Signal Efficiency}
\label{sec:CutsAndEfficiency}
\subsection{Data Selection}
\label{sec:DataSelection}

Once the first level data processing is complete, we select the events
of interest with a set of event cuts. These cuts fall into three
types:
\begin{enumerate}
\item Time-based cuts that remove periods of time where the data
  quality were poor or the data processing failed.
  \label{item:timebased}
\item Event-based cuts that remove poorly reconstructed and
  non-signal-like events to maximize sensitivity to {\BBless} decay.
  \label{item:eventbased}
\item Anti-coincidence cuts that remove events that occur in multiple
  bolometers and are thus less likely to come from a {\BBless} decay.
  \label{item:anticoincidence}
\end{enumerate}

\textbf{\ref{item:timebased}. Time-Based Selection:} The first set of
cuts removes intervals of time where the data collection was
poor. This typically removes periods of excessive noise from an
individual bolometer (e.g. a malfunctioning electronic channel), or
periods of time when the entire detector temperature was fluctuating
quickly (e.g. during an earthquake). This cut introduces a dead time
that reduces our total exposure by 3.5\%. We further remove intervals
of time when the data processing failed. The most significant
component of this was a failure of the TGS algorithms to stabilize
gain variations over too large a temperature drift. These excluded
periods lead to the reduction in our total exposure of 2.7\%
mentioned in the previous section.

\textbf{\ref{item:eventbased}. Event-Based Selection:} We implement a
set of event based cuts that remove events that are either
non-signal-like or are in some way not handled well by the data
processing software. This includes a set of basic quality cuts that
removes events that are clearly problematic, such as events that
exceed the dynamic range of the electronics or events that overlap
with one of the injected heater pulses. We further implement a pile-up
cut that rejects an event if more than one trigger occurs in the same
bolometer within 3.1\,s before or 4\,s after the event trigger. This
7.1\,s window allows any previous event enough time to return to
baseline and ensures that any following event does not occur within
the event window.

In addition to these basic quality checks, we have developed a set of
pulse shape cuts, which remove events on the basis of six pulse shape
parameters. These include the slope of the baseline as well as the
time in the event window that the signal reaches its maximum. Cutting
on these two parameters is useful for removing events whose amplitudes
are poorly reconstructed by the processing software. The pulse shape
cuts also cut on the pulse rise and decay times, which are useful for
identifying pile-up events that failed to cause a second trigger and
events that have very fast time constants and are believed to be due
to energy depositions in the thermistor itself, fast temperature
variations due to vibrations, or electronic noise. The last pulse
shape cut selects on two parameters referred to as ``Test Value Left''
(TVL) and ``Test Value Right'' (TVR). These are effectively $\chi^2$
values between the normalized OF filtered pulse shape and the expected
filtered detector response shape on either the left or right side of
the signal peak. These last two parameters are useful for identifying
events whose shape deviates significantly from the expected detector
response shape.

All pulse shape parameters have an energy dependence, which we
normalize by interpolating across events that lie within peaks in the
calibration spectrum over the range 146\,keV to 2615\,keV. As a
result, the efficiency of the cuts on these variables is independent
of energy to within statistical uncertainty over this range. We tune
these pulse shape cuts by maximizing the signal efficiency over the
square root of the background in the physics spectrum, where the
signal efficiency is measured as the fraction of selected events in
the $\gamma$ peaks over the range 146 -- 2615\,keV, and the
background is measured in the energy regions around the peaks. To
avoid biasing our selection, we use a randomly selected half of the
data for tuning the selection and the remaining events for determining
the selection efficiency (reported in
Table~\ref{tab:SignalEfficiency}). We exclude the {\BBless} decay ROI
from both calculations.

\begin{figure}
  \centering
  \includegraphics[width=.98\columnwidth]{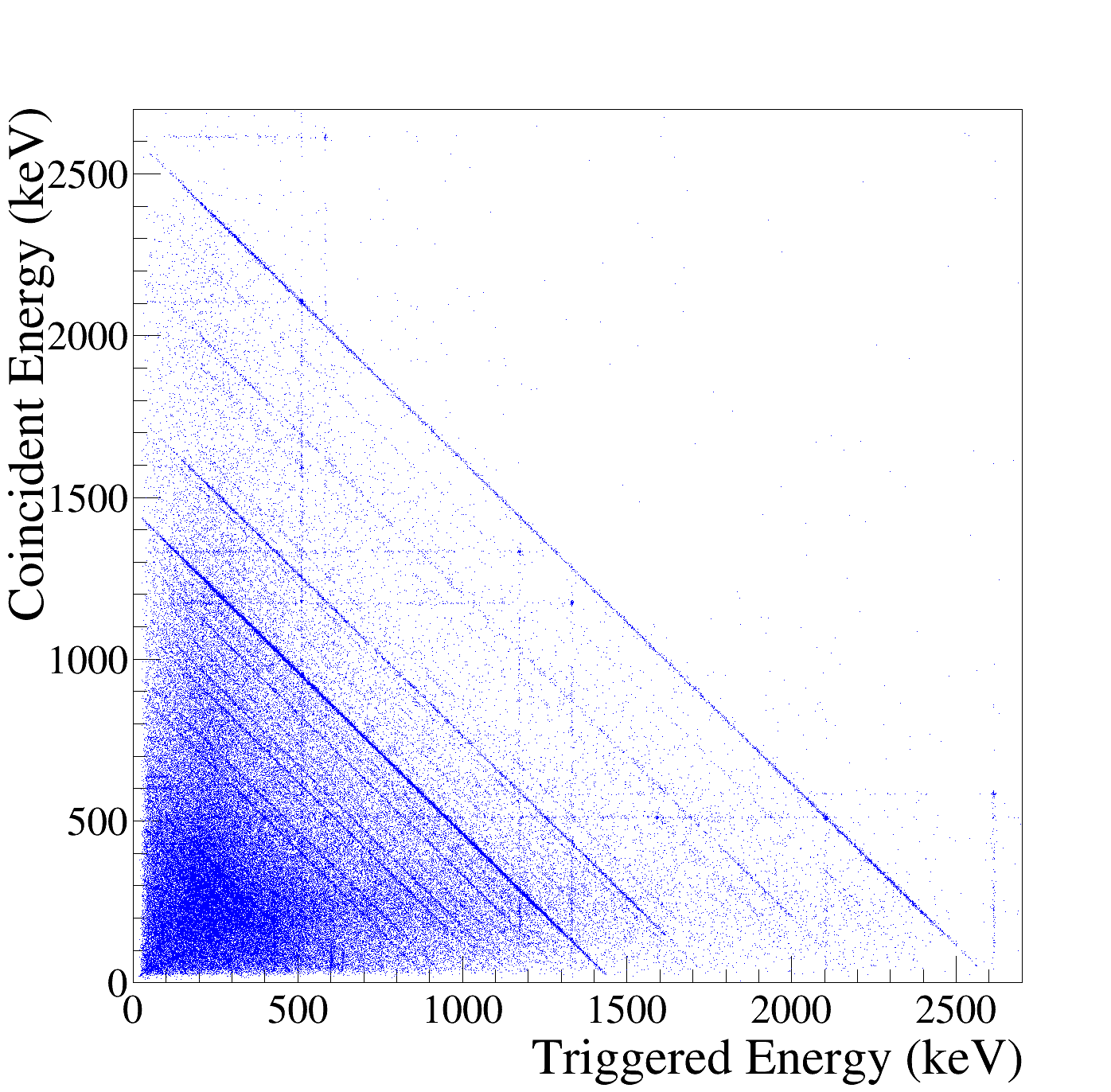}
  \caption{Plot of the two absorbed energies in double-crystal
    coincidences during physics data collection. The diagonal lines
    correspond to events where a $\gamma$ scatters in one crystal and
    is then fully absorbed in another. The vertical and horizontal
    lines are cascade events where one $\gamma$ is fully absorbed and
    the other is scattered. This can be seen for the two $^{60}$Co
    $\gamma$-rays, $^{208}$Tl 2615\,keV + 583\,keV $\gamma$-rays, and
    $^{208}$Tl pair production events where one annihilation photon
    escapes and is absorbed in another bolometer.}
  \label{fig:E1_vs_E2}
\end{figure}

\begin{figure}[h!]
  \centering
  \includegraphics[width=.98\columnwidth]{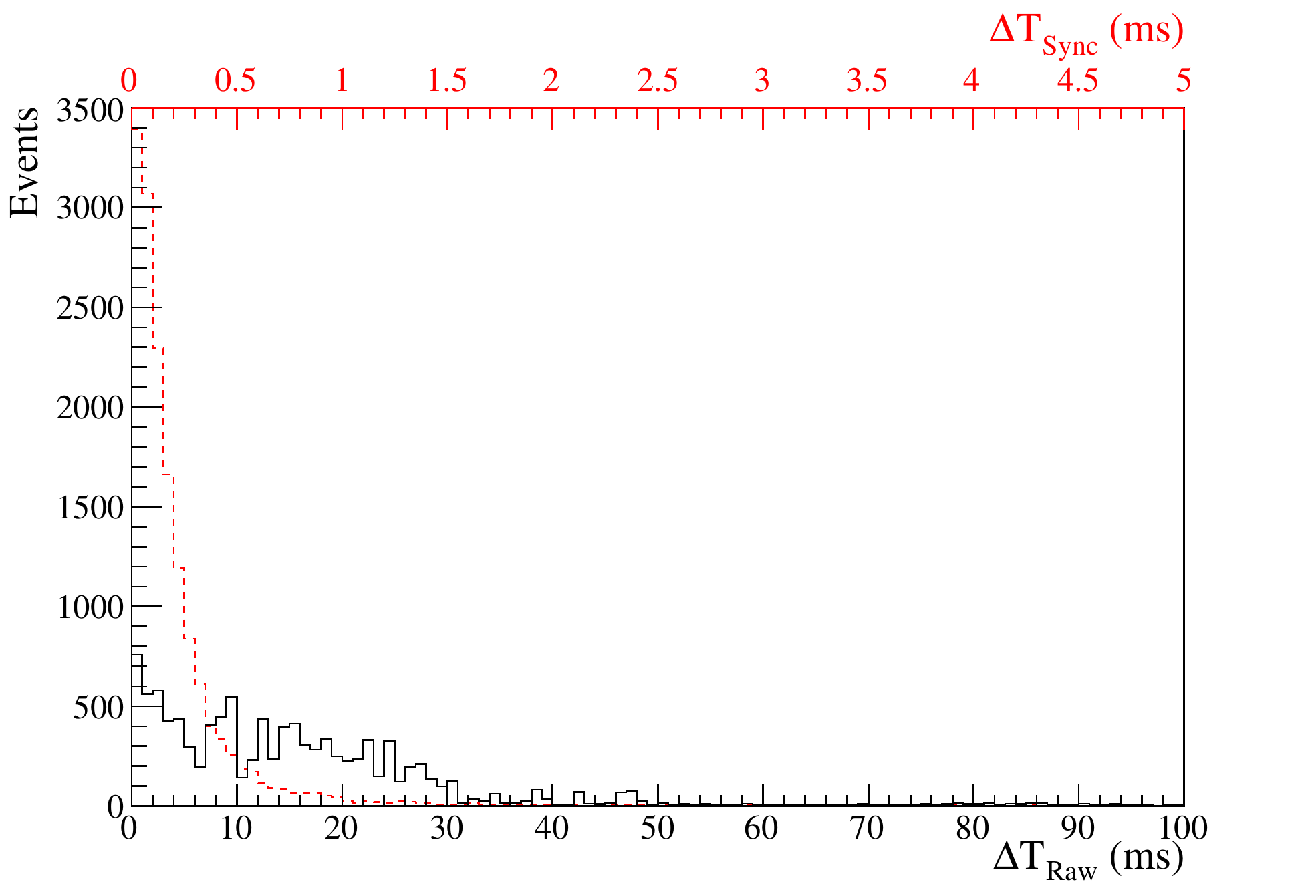}
  \caption{Distribution of measured time between coincident events
    before correcting for different detectors' rise time (\emph{black
      solid}, $\Delta T_{\rm Raw}$) and after
    (\emph{\grayselect{gray}{red dashed}}, $\Delta T_{\rm Sync}$).}
  \label{fig:JitterPlots}
\end{figure}

\textbf{\ref{item:anticoincidence}. Anti-coincidence Selection:} Since
the desired {\BBless} decay events have their full energy absorbed in
a single bolometer, we use an anti-coincidence cut to reject any event
that occurs within $\pm$5\,ms of another event in any other bolometer
in the tower. This cut primarily rejects $\alpha$-decays that occur on
the surfaces of our bolometers, $\gamma$-rays that scatter in one
bolometer before being absorbed in another, cascade $\gamma$-rays from
radioactive decays, and muons passing through the tower and their
secondary neutrons. A plot of the energies of double-crystal
coincidence events --- events where two bolometers are triggered ---
is shown in Fig.~\ref{fig:E1_vs_E2}. In \QINOPaper, the
anti-coincidence window was $\pm$50\,ms, and in~\cite{Q0InitialPaper}
we used a window of $\pm$100\,ms. Here we have significantly narrowed
this window by accounting for the constant differences in detector
rise times between different bolometers when measuring the time
between two events on different bolometers (see
Fig.~\ref{fig:JitterPlots}). This correction improves the timing
resolution by a factor of $\approx50$. 

After implementing all cuts, 233 out of 411 triggered events remain in
the ROI for the {\BBless} analysis described in
Sec.~\ref{sec:NDBDAnalysis}.

\subsection{Signal Detection Efficiency}
\label{sec:Efficiency}
The signal detection efficiency for a {\BBless} decay is a product of
conditional probabilities: the probability that the full energy of the
decay is contained in a single crystal, the probability that the event
is then triggered and properly reconstructed, the probability that the
event then passes the signal cuts, and the probability that the event
is not then accidentally in coincidence with an unrelated event in a
different bolometer. These efficiencies are summarized in
Table~\ref{tab:SignalEfficiency}.

We use a {\sc Geant}4-based~\cite{geant4} Monte Carlo simulation to
estimate the fraction of events that deposit their full energy in a
single crystal. This simulation models the most significant energy
loss mechanisms: electron escape, X-ray escape, and the escape of
Bremsstrahlung photons. The simulation also mimics the detector
response by convolving the spectrum with a Gaussian to reproduce the
expected shape near $Q_{\beta\beta}$. We calculate the efficiency by
fitting the resulting {\BBless} decay peak and dividing the fitted
area by the number of simulated decays. The efficiency evaluates to
\mbox{$88.345\pm0.040({\rm stat})\pm0.075(\rm syst)$\%}. The
systematic uncertainty is from the variation in the crystal
dimensions, the uncertainty in decay energy, and the step choice for
secondary propagation in the {\sc Geant}4 simulation.

We evaluate the trigger and energy reconstruction efficiencies using
the pulser heater events. The DAQ automatically flags each heater
event in the data, and then passes the event through the standard
signal trigger algorithm. The fraction of heater events that also
generate a signal trigger provides an estimate of our signal trigger
efficiency. The heater events typically reconstruct as a Gaussian peak
around $3-3.5$\,MeV. We determine our energy reconstruction efficiency
by fitting this heater peak with a Gaussian line shape and counting
the fraction of events that reconstruct within 3$\sigma$. This
calculation is done for each bolometer for each physics run and
averaged, weighted by exposure, to determine a single efficiency for
the entire detector. \edit{The bolometers without working heaters are
  excluded from this calculation and are assigned the same efficiency
  as the other bolometers --- thus they are assumed to have the
  average efficiency of the other bolometers.}

We estimate the efficiency of the signal cuts (i.e., pile-up and pulse
shape) using the $^{208}$Tl 2615\,keV peak in the physics data. The
vast majority of events that reconstruct in the peak are properly
reconstructed; pile-up events and events with non-standard pulse
shapes reconstruct somewhat randomly with a much wider
distribution. We estimate our cut efficiency by measuring the rate
within 5$\sigma$ of the 2615\,keV peak and subtracting the background
rate measured in bands around the peak. We compare this signal rate
before and after applying the cuts to determine the fraction of signal
events that are accidentally removed by the signal cuts.

The anti-coincidence efficiency accounts for the rejection of valid
events due to an event being close enough in time to an unrelated
event on another bolometer so as to accidentally be considered a
coincidence. This efficiency is estimated in a similar fashion to the
signal cut efficiencies --- comparing the signal rate before and after
the cut --- except that it is calculated around the 1460\,keV line from
electron capture in $^{40}$K. While the $^{208}$Tl 2615\,keV
$\gamma$-ray can be part of cascade and is expected to occasionally
occur in coincidence with other $\gamma$-rays, the $^{40}$K 1460\,keV
only occurs in coincidence with a 3\,keV X-ray which is well below the
trigger threshold of our bolometers. Thus any event in coincidence
with a fully absorbed 1460\,keV $\gamma$-ray constitutes an accidental
coincidence.

Combining these, we determine the total signal efficiency of the {\q}
detector to be $81.3\pm0.6$\%.

\begin{table}[h!]
  \centering
  \caption{{\q} signal detection efficiency. See the text for how
    these are calculated.}
  \begin{tabular}{cl}
    \hline
    \hline
    Source & \multicolumn{1}{c}{Signal Efficiency (\%)} \\
    \hline
     {\BBless} energy confinement & $88.345\pm0.040({\rm stat})\pm0.075({\rm syst})$ \\
    Trigger \& Reconstruction     & $98.529\pm0.004$\\
    Pile-up \& Pulse Shape cuts   & $93.7\pm0.7$\\
    Anti-coincidence cut          & $99.6\pm0.1$\\
    \hline 
    Total & $81.3\pm0.6$\\
    \hline
    \hline
  \end{tabular}
  \label{tab:SignalEfficiency}
\end{table}

 \begin{figure*}
  \centering
  \includegraphics[width=\textwidth]{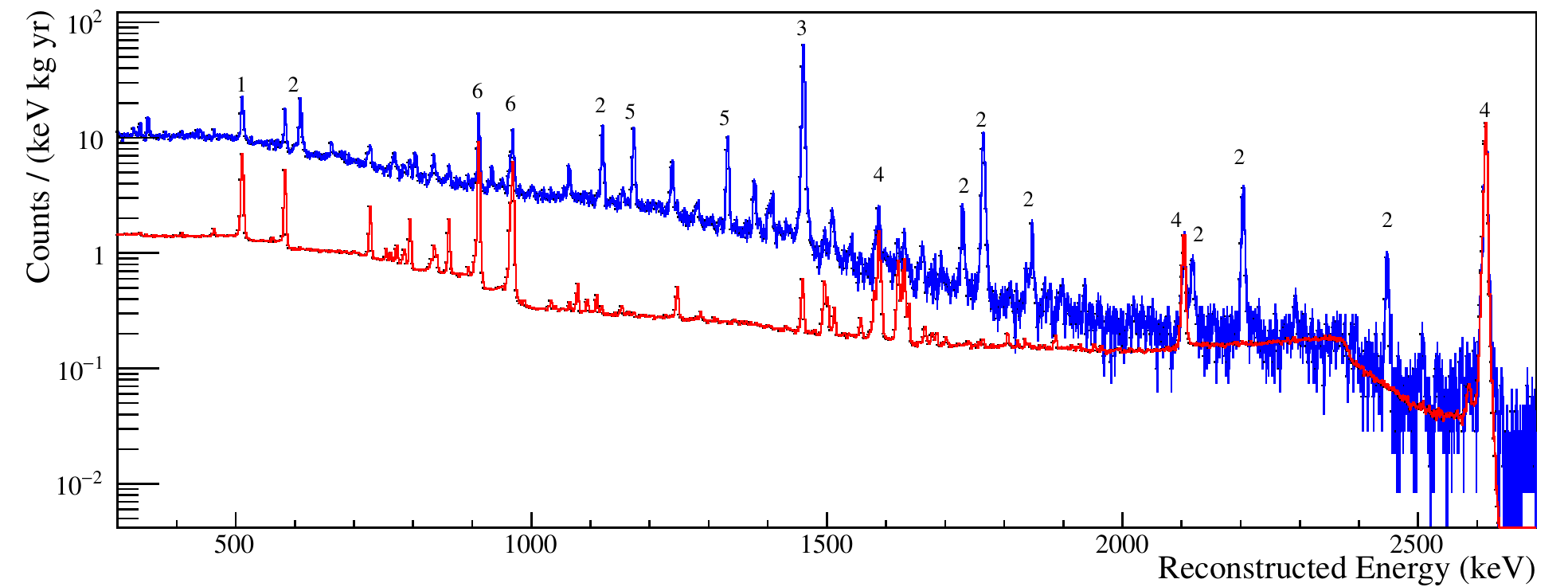}

  \caption{The final {\q} physics spectrum
    (\emph{\grayselect{black}{blue}}) and calibration spectrum
    (\emph{\grayselect{gray}{red}}). The calibration spectrum has been
    normalized to match the rate of the physics spectrum around the
    2615~keV $^{208}$Tl peak. The most prominent peaks in the physics
    spectrum are from the decay of known radioactive backgrounds: (1)
    $e^+e^-$ annihilation, (2) $^{214}$Bi, (3) $^{40}$K, (4)
    $^{208}$Tl, (5) $^{60}$Co and (6) $^{228}$Ac. Figure adapted from
    \cite{Q0FinalPrl}.}
  \label{fig:LabeledSpectrum}
\end{figure*}

\section{Data Unblinding}
\label{sec:DataUnblinding}
\begin{figure*}
  \centering
  \subfigure{
    \includegraphics[width=.48\textwidth]{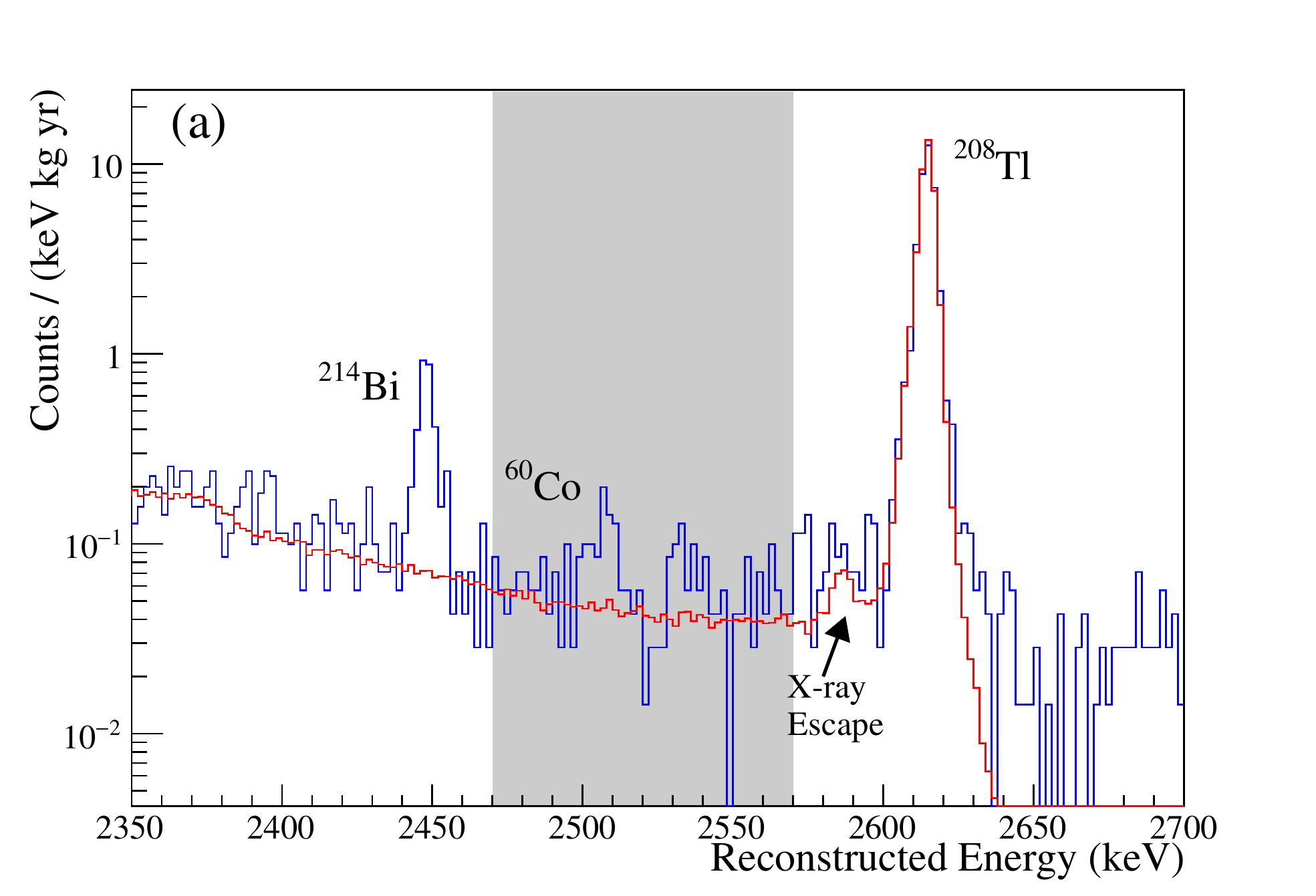}
    \label{fig:ExpandedROISpectrum}
  }
  \subfigure{
    \includegraphics[width=.48\textwidth]{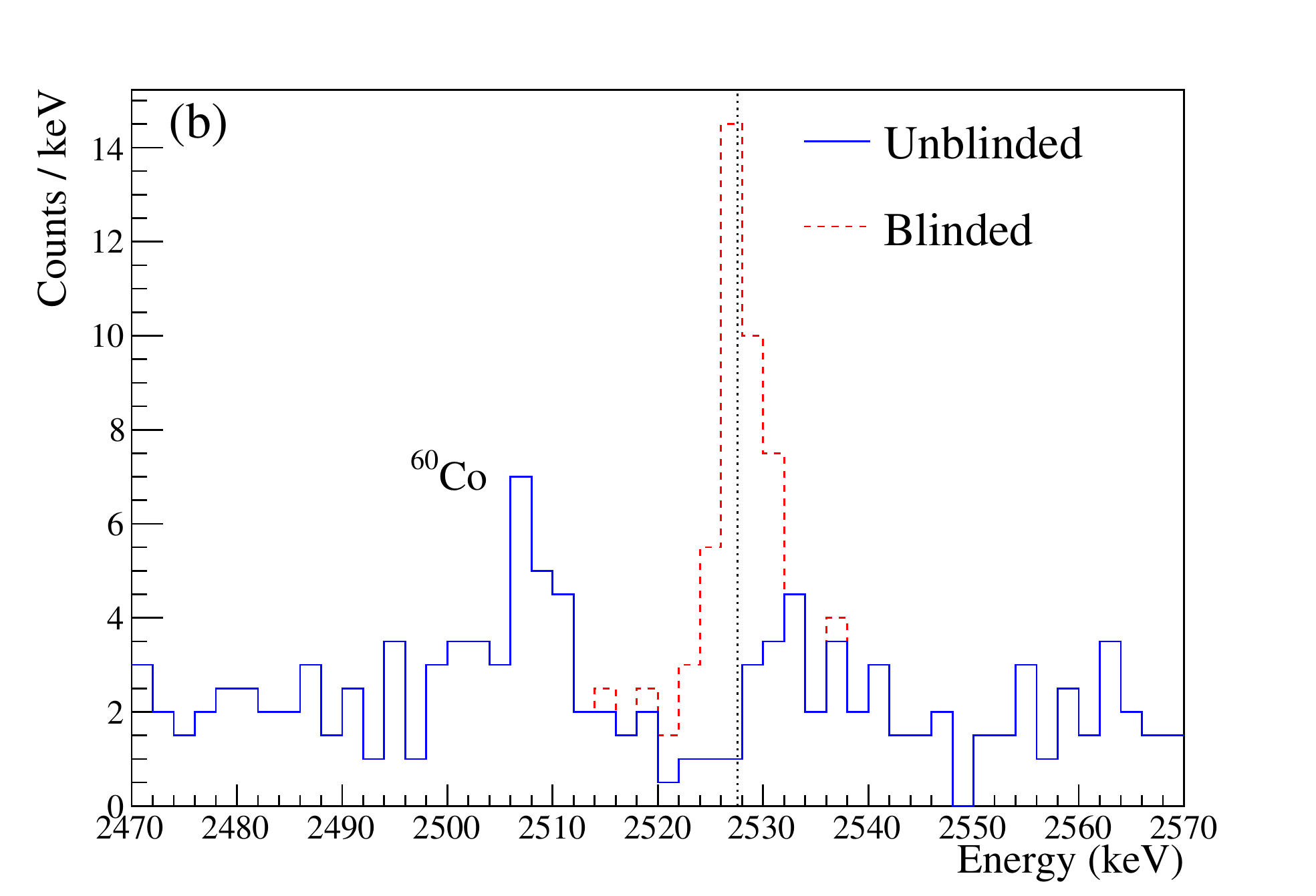}
    \label{fig:BlindedUnblinded}
  }
  \caption{(a) The {\q} spectrum around the ROI. This is a zoomed view
    of Fig.~\ref{fig:LabeledSpectrum}. The shaded region corresponds
    to the energy range used in the ROI fit. The background in the ROI
    is now dominated by the scattered-$\gamma$ background rather than
    the flat $\alpha$ background.  (b) Comparison of the blinded
    (\emph{dashed}) and unblinded (\emph{solid}) spectra in the
    ROI. The peak in the dashed spectrum is the artificial peak
    created by the blinding procedure. $Q_{\beta\beta}$ is indicated
    by the dotted line.}
\end{figure*}

The unblinding procedure was decided upon before any data were
unblinded. After fixing the data selection cuts and the {\BBless}
decay analysis procedure (described in the next section), we unblinded
the data in two stages: first we unblinded 17 of 20 datasets (or
8\,{\kgyr} of $^{130}$Te exposure) and began the {\BBless} decay
analysis while we continued to collect the final three datasets (or
1.8\,{\kgyr} of $^{130}$Te exposure). The last three datasets were
blinded during collection, were subjected to the same production
procedure and cuts as the rest of the data, and were to be included
regardless of their effect on the final result. The unblinded spectrum
is shown in Fig.~\ref{fig:LabeledSpectrum}, and the blinded and
unblinded spectra in the ROI are shown in
Fig.~\ref{fig:BlindedUnblinded}.

As a cross-check, we also reproduced all of the {\q} data without the
blinding/unblinding steps and compared them to the data that had been
blinded and unblinded to confirm it had no effect on the final
spectrum. Indeed, the blinding/unblinding procedure had no effect on
our final spectrum. This confirmation of our blinding/unblinding
procedure validates this technique moving forward to \Q.

\section{{\BBless} Analysis}
\label{sec:NDBDAnalysis}

The {\q} physics spectrum over the range 300--2700\,keV is shown in
Fig.~\ref{fig:LabeledSpectrum}. The {\q} data consists of 20 datasets
collected on 51 active bolometers. After implementing all cuts, 1,008
bolometer-dataset (\BoDs) pairs remain for a total TeO$_2$ exposure of
35.2\,{\kgyr}, or 9.8\,{\kgyr} of $^{130}$Te. Our {\BBless} decay
analysis treats each one of these as a semi-independent experiment
with some parameters unique to each {\BoDs}, some parameters shared
across datasets (i.e., constant in time), and other parameters shared
globally (i.e., constant in time and uniform across the detector).

We define the ROI for our {\BBless} decay analysis as the range
2470--2570\,keV; this region contains 233 events. This is the widest
possible range that allows us to constrain the background rate without
introducing unnecessary peaks into the analysis.  The range is bounded
by a $^{214}$Bi line at 2448\,keV and a small peak at 2585\,keV from a
2615\,keV $^{208}$Tl $\gamma$-ray minus a 30\,keV Te X-ray escape (see
Fig.~\ref{fig:ExpandedROISpectrum}). The ROI contains the potential
{\BBless} decay signal at 2527\,keV as well as a peak from the
single-crystal coincidence of the two $\gamma$-rays from $^{60}$Co
decay which lies only 21\,keV below. We attribute this $^{60}$Co
contamination to the activation of the copper frames and internal
shielding~\cite{Q0BackgroundModel}. We have measured the production
rate of $^{60}$Co inside the TeO$_2$ crystals to be
small~\cite{Wang2015}, so we expect a negligible background from the
$\beta+\gamma+\gamma$ coincidence.

Our {\BBless} decay analysis proceeds through three steps. We first
construct a detector response function $\rho_\bd$ for each {\BoDs},
which characterizes the expected spectral shape of a particular
bolometer's response to a mono-energetic energy deposition during a
particular dataset. We then use this set of $\rho_\bd$ to fit other
prominent peaks in the physics spectrum. This allows us to understand
how our detector response depends on energy. Finally, we fit the ROI
by postulating a peak at the {\BBless} decay energy and constraining
its amplitude with a detector response function properly scaled in
energy. The resulting best-fit amplitude provides insight into the
{\BBless} decay rate.

\subsection{Detector Energy Response}
\label{sec:Lineshape}
We model the detector response to the mono-energetic {\BBless} decay
signal based on the measured response to the $\gamma$ peaks. This is
done for each {\BoDs}, $i$, using the functional form
\begin{equation}
\begin{array}{r}
\rho_\bd(E;\mu_\bd,\sigma_\bd,\delta_\bd,\eta_\bd)\equiv\hspace{1cm}\;\\
(1-\eta_\bd){\rm
  Gauss}(E;\mu_\bd,\sigma_\bd)\\+\eta_\bd{\rm
  Gauss}(E;\delta_\bd\mu_\bd,\sigma_\bd).
\end{array}
\end{equation}
This function produces a primary Gaussian centered at $\mu_\bd$ and a
secondary Gaussian at a slightly lower energy $\delta_\bd\mu_\bd$,
with $\delta_\bd\sim0.997$. This smaller secondary peak accounts for
$\eta_\bd\sim5\%$ of events and models an energy loss mechanism whose
origin is presently under investigation. The presence of this
substructure is unaffected by the choice of pulse filtering technique
or TGS algorithm and is present on all channels. It is not clustered
in time \edit{or a result of pile-up of events. It also does not
  appear to be correlated with any shape parameter used in the above
  cuts. A visual inspection of pulses selected from the primary and
  secondary peaks reveals no obvious difference in the pulse
  shape.} The {\qino} data shows a hint of this asymmetric line shape;
however, it is the improved resolution of the {\q} detector that has
made this effect clear. We tested multiple models to reproduce the
line shape, including a Gaussian distribution with an asymmetric tail
and a triple Gaussian lineshape which modeled escapes of 4\,keV
characteristic X-rays from Te. Ultimately, we settled on the
double-Gaussian shape which reproduced the data well across a broad
range of energies.

Each {\BoDs} has its own peak position, $\mu_\bd$, and a single
resolution parameter, $\sigma_\bd$, for both the primary and secondary
Gaussian peaks. The data suggest that the position and amplitude of
the secondary Gaussian peak may vary between bolometers and in time,
thus indicating that this is possibly a detector related effect. Both
$\delta_\bd$ and $\eta_\bd$ are free to vary from bolometer to
bolometer, but to limit the number of free parameters both are
constant in time within each of the two data-taking campaigns.

\begin{figure*}
  \centering
  \subfigure{
    \includegraphics[width=.48\textwidth]{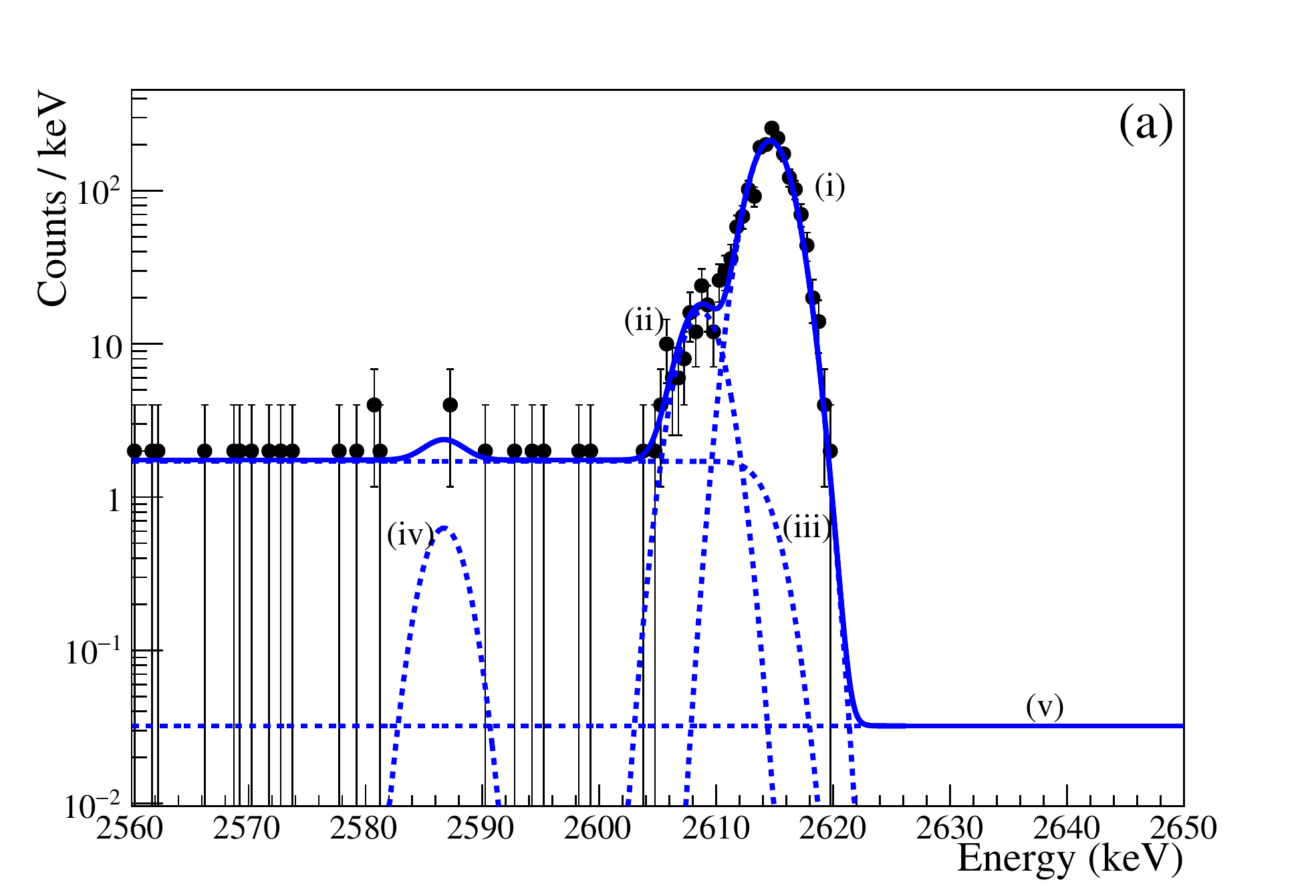}\hfill
    \label{fig:SingleChannelCalibrationFit}
  }
  \subfigure{
    \includegraphics[width=.48\textwidth]{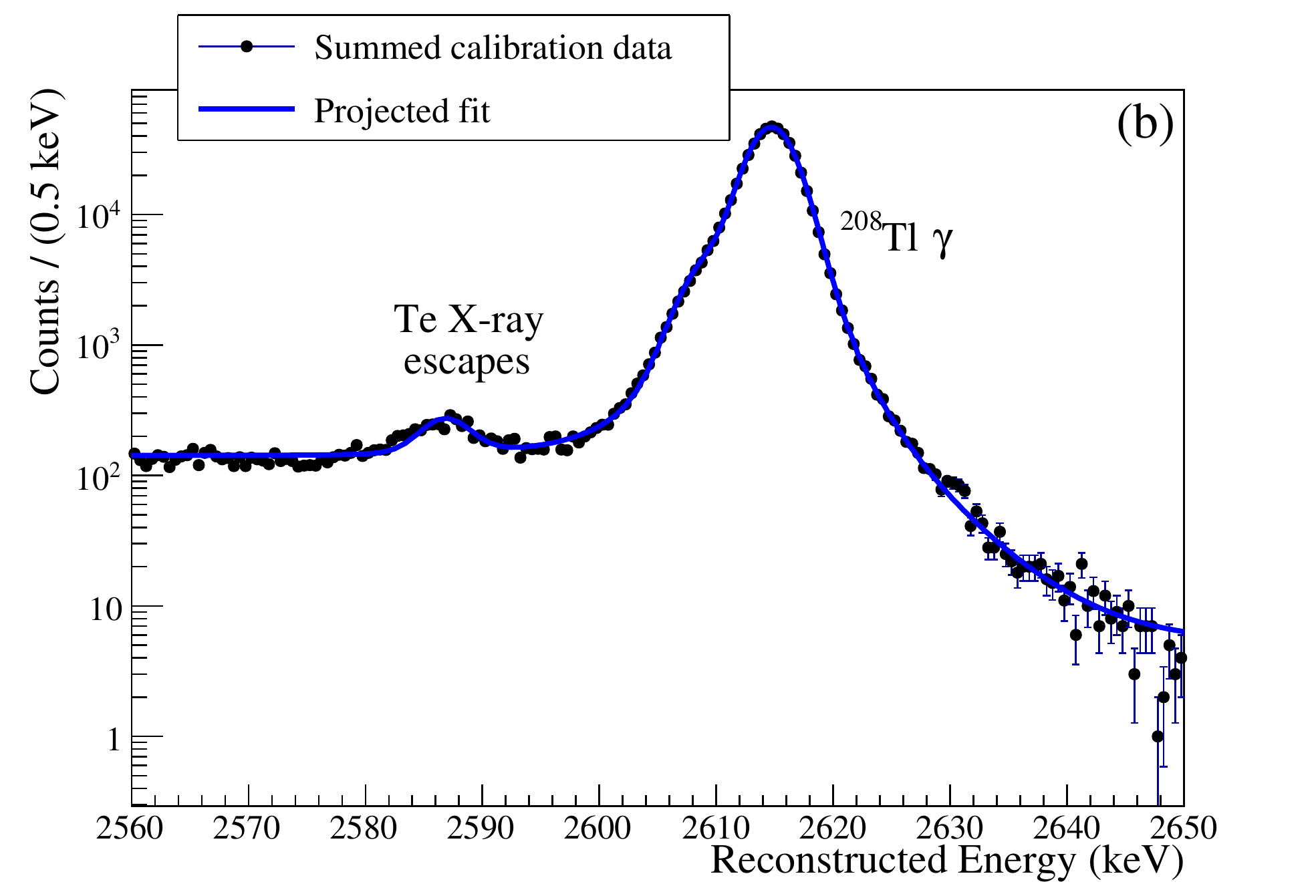}
    \label{fig:SummedChannelCalibrationFit}
  }
  \caption{(a) Best-fit model to the calibration 2615\,keV peak for a
    single bolometer-dataset. The solid line shows the full summed
    model, and the dashed lines show the individual components: (i)
    the primary peak of the detector response; (ii) the secondary peak
    of the detector response; (iii) the smeared step-function; (iv)
    the X-ray escape peak; (v) flat background. (b) The calibration
    spectrum summed over all {\BoDs} pairs, with the summed best-fit
    model. Figure adapted from \cite{Q0FinalPrl}.}
\end{figure*}

We estimate the best-fit detector response for each {\BoDs} by fitting
the intense $^{208}$Tl 2615\,keV calibration peak. This fit is over
the range 2560--2650\,keV and includes three more elements to model
the background under the detector response: (i) a smeared step
function, modeled as an Erfc function, to model $\gamma$-rays that
scatter in the shields before interacting with a bolometer or scatter
multiple times in a single bolometer before exiting; (ii) a Gaussian
peak roughly 30\,keV below the primary peak to model an event in which
a 2615\,keV $\gamma$-ray is absorbed and one of the characteristic Te
K shell X-rays, which have energies that range from 27--31\,keV, is
produced and escapes the crystal; (iii) a flat background. The best
fit for a single {\BoDs} is shown in
Fig.~\ref{fig:SingleChannelCalibrationFit}.

The full calibration peak model is given by
\begin{equation}
  \begin{array}{rcl}
    f^{\rm Tl}_\bd(E)&=&
    R^{\rm Tl}_\bd\rho_\bd(E;\mu_\bd,\sigma_\bd,\delta_\bd,\eta_\bd)\\ 
  &&+r_{\rm Scatter}R^{\rm Tl}_\bd{\rm Erfc}(\frac{E-\mu_\bd}{\sqrt{2}\sigma_\bd})\\ 
    &&+r_{\rm Escape}R^{\rm Tl}_\bd{\rm Gauss}(E;\delta_{\rm Escape}\mu_\bd,\sigma_\bd)\\ 
    &&+b^{\rm Cal},
  \end{array}
\end{equation}
where $R^{\rm Tl}_\bd$ represents a {\BoDs} dependent $^{208}$Tl peak
event rate in {\cky}, which is a free parameter in the fit. The event
rates of both the scattered $\gamma$-rays and the X-ray escape peak
are given as fractions of the peak event rate, $r_{\rm Scatter}$ and
$r_{\rm Escape}$ respectively. Both of these are global physical
parameters that could be estimated using Monte Carlo, but since
modeling them requires carefully accounting for detector thresholds
(to accurately predict the fraction that are flagged as a coincidence)
these parameters are instead left unconstrained in the fit. The
position of the X-ray escape peak is described as a fraction of the
primary peak energy, $\delta_{\rm Escape}$, and is also left
unconstrained in the fit. The final parameter $b^{\rm Cal}$ is a
global flat background rate in \ckky, also unconstrained.

We perform a simultaneous unbinned extended maximum likelihood (UEML)
fit to all {\BoDs} pairs using the \textsc{RooFit} fitting
package~\cite{RooFit}. For each {\BoDs}, this yields a set of
parameters which describe the detector response function,
$(\hat\mu_\bd,\hat\sigma_\bd,\hat\delta_\bd,\hat\eta_\bd)$. We fix
these parameters for use later in the ROI
fit. Figure~\ref{fig:SummedChannelCalibrationFit} shows the resulting
best-fit model to the summed calibration data over all {\BoDs} pairs.

\subsection{Detector Response as a Function of Energy}
\label{sec:LineshapeVsEnergy}
For each {\BoDs}, we consider the parameters that characterize
the detector response at 2615\,keV,
$(\hat\mu_\bd,\hat\sigma_\bd,\hat\delta_\bd,\hat\eta_\bd)$, as
fixed. By fitting this detector response to other $\gamma$ lines in
the physics spectrum, we can derive the energy dependence of the
detector response, in particular the predicted response at
$Q_{\beta\beta}$. Specifically, we seek to account for:
\begin{itemize}
\item any bias in the reconstructed energy of a {\BBless} decay
  signal;
\item the dependence on energy of the detector energy resolution.
\end{itemize}
Since the 2615\,keV peak lies only 87\,keV above $Q_{\beta\beta}$, we
use it as an anchor for our ROI fit. Most importantly, for each
{\BoDs} we use the reconstructed energy of the $^{208}$Tl peak in the
calibration data to dictate the energy at which we search for a
{\BBless} decay peak, i.e., rather than fixing our fit position to
$Q_{\beta\beta}$, we search for a peak 87\,keV below $\hat\mu_\bd$ in
each {\BoDs}; this allows us to decrease the impact of any residual
miscalibration. For this, we analyze the prominent $\gamma$ lines in
the physics spectrum over the energy range 300--2500\,keV.

\begin{figure}
  \centering
  \subfigure{
    \includegraphics[width=.95\columnwidth]{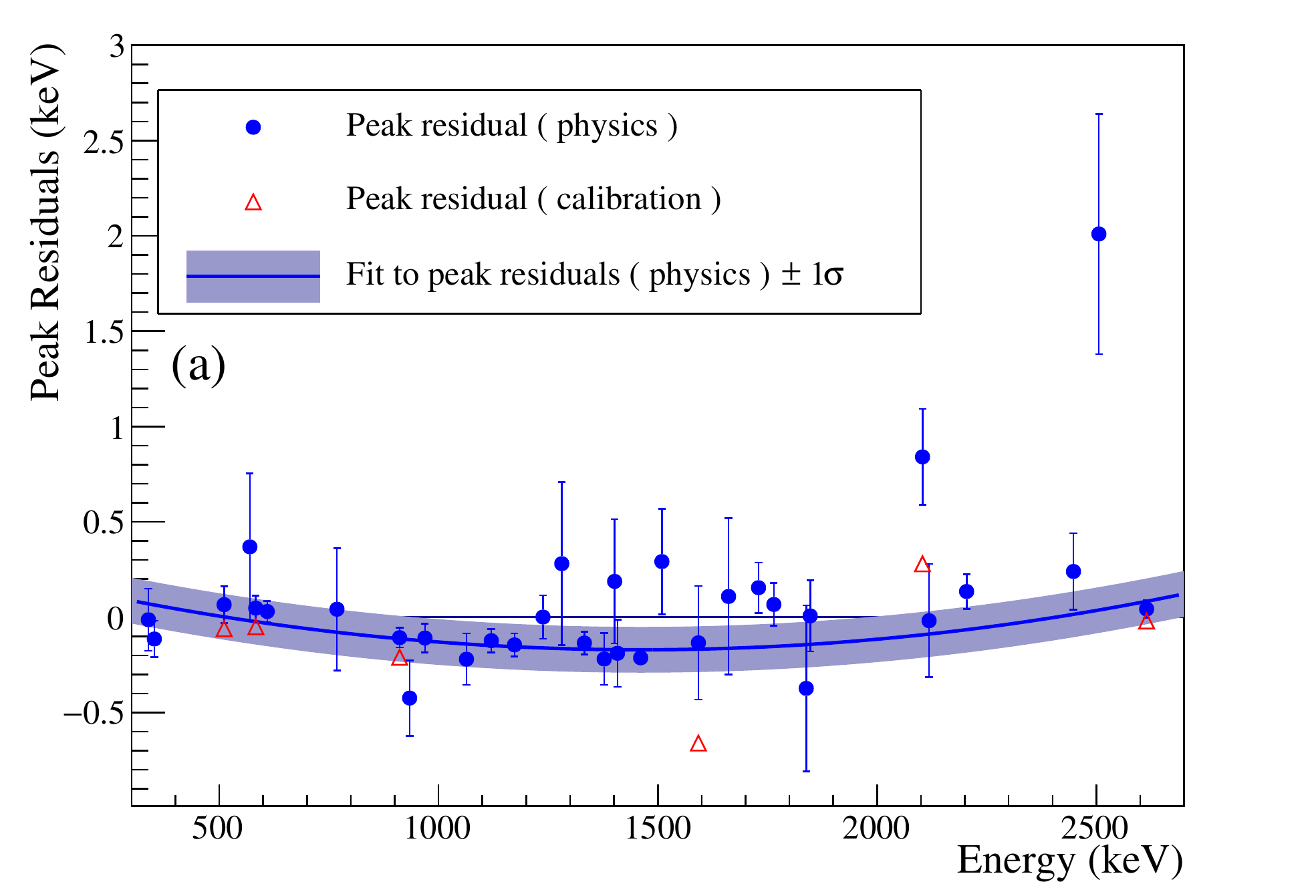}
    \label{fig:RelativeResiduals}
  }
  \subfigure{
    \includegraphics[width=.95\columnwidth]{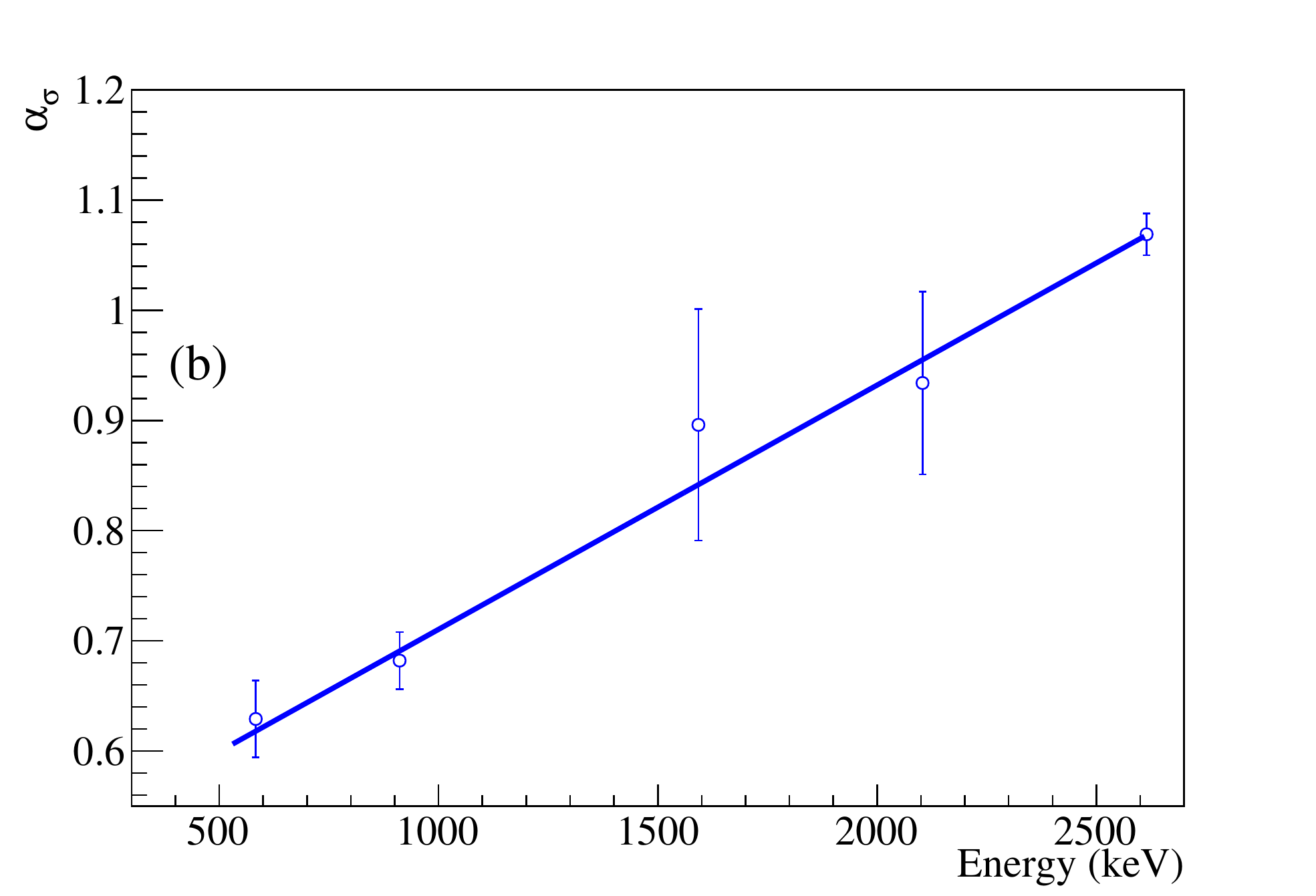}
    \label{fig:ResolutionScaling}
  }
  \caption{(a) Residuals of the best-fit reconstructed peak
    energy and expected peak energy (fit - expected) for the physics
    data (\emph{\grayselect{black}{blue} circles}) and calibration
    data (\emph{\grayselect{gray}{red} triangles})\edit{ --- the
      statistical errors on the calibration points are
      negligible}. The \grayselect{black}{blue} curve and shaded band
    are the fit to the physics peak residuals $\Delta\mu(E)$ and the
    1$\sigma$ uncertainty band. Figure adapted from
    \cite{Q0FinalPrl}. (b) Best-fit resolution scaling
    parameter, $\alpha_\sigma$, for a few of the peaks in the physics
    spectrum, as well as the best-fit interpolation.}
\end{figure}

Anticipating the approach to fitting the ROI described in the next
section, we take an analogous approach to fitting the $\gamma$ peaks
in the physics spectrum. For each $\gamma$ line, we perform an UEML
fit with the position of the detector response shifted down from the
$^{208}$Tl peak position, $\hat\mu_\bd\rightarrow\hat\mu_\bd-\Delta$,
and all of the energy resolutions scaled by a fixed amount,
$\hat\sigma_\bd\rightarrow\alpha_\sigma\hat\sigma_\bd$. The shift in
energy $\Delta$ \edit{parameterizes the difference in energy between
  the reconstructed peak in the physics spectrum and the reconstructed
  $^{208}$Tl peak in the calibration spectrum. The scaling of the
  energy resolution $\alpha_\sigma$ parameterizes the resolution
  scaling both as a function of energy and between the calibration and
  physics data.} In the fit, both parameters are unconstrained and the
same for all {\BoDs} pairs. The parameters of the secondary peak,
$\delta_\bd$ and $\eta_\bd$, are held fixed. Each fit also includes a
background model that is either a 1st or 2nd-degree
polynomial. Including $\Delta$ and $\alpha_\sigma$, each fit has four
to five free parameters.

For each peak, we compare the best-fit shift from the $^{208}$Tl
calibration peak to the expected shift for that peak to determine its
energy reconstruction residual $\Delta\mu$. The residuals for 33
prominent peaks in the physics spectrum and 6 peaks in the calibration
spectrum are shown in Fig.~\ref{fig:RelativeResiduals}. Note that
these residuals are relative to the 2615\,keV peak in the calibration
data, thus include any residual miscalibration both as a function of
energy and between the calibration and physics spectra. The residuals
display a parabolic energy dependence, which we attribute to a
systematic bias in the energy calibration step of the data
processing. 

We account for this systematic miscalibration by shifting the position
at which we expect our {\BBless} decay signal to occur. We fit a
second order polynomial to the peak residuals, $\Delta\mu(E)$, and
evaluate the expected residual at the $^{130}$Te Q-value, and use the
weighted RMS of the residuals about $\Delta\mu(E)$ as the systematic
uncertainty, \mbox{$\Delta\mu(Q_{\beta\beta})=0.05\pm0.05\,{\rm
    (stat)}\pm0.09\,{\rm (syst)}$\,keV}. As is evident from
Fig.~\ref{fig:RelativeResiduals} the $^{60}$Co
single-crystal coincidence peak shows a higher than expected residual
and so \edit{this peak as well as the $^{208}$Tl single-escape peak at
  2103\,keV are excluded from the evaluation of $\Delta \mu(E)$;} we
comment further on these peaks in Section~\ref{sec:DetectorPerformance}.

We perform a similar interpolation to estimate the energy resolution
scaling at $Q_{\beta\beta}$, which evaluates to
\mbox{$\alpha_\sigma(Q_{\beta\beta})=1.05\pm0.05$} (see
Fig.~\ref{fig:ResolutionScaling}). \edit{As above, this
  scaling includes both an energy dependent component as well as any
  bias between the calibration and physics data.} The uncertainty on
this scaling is purely systematic, and driven by the choice of
$\gamma$-lines to include in the fit. For the central value quoted
here, we include only the $\gamma$-peaks that are well defined in both
the physics and calibration spectra.

\subsection{Fitting the Region of Interest}
\label{sec:ROIFit}
The fit to the ROI follows an analogous process to the other peaks in
the physics spectrum. We simultaneously fit both the hypothetical
{\BBless} decay signal and the peak from the single-crystal $^{60}$Co
coincidence. For each {\BoDs}, we model the ROI as
\begin{equation}
  \begin{array}{rcl}
    f_\bd(E)&=&R^{0\nu}\rho_\bd(E;E_\bd^{0\nu},\alpha_\sigma(Q_{\beta\beta})\hat\sigma_\bd,\hat\delta_\bd,\hat\eta_\bd) \\
    &&+R^{\rm Co}(t)\rho_\bd(E;E_\bd^{\rm Co},\alpha_\sigma(Q_{\beta\beta})\hat\sigma_\bd,\hat\delta_\bd,\hat\eta_\bd)\\
    &&+b_{\rm ROI},
  \end{array}
\end{equation}
where $R^{\rm Co}(t)$ and $R^{0\nu}$ are the event rates in {\cky} for
$^{60}$Co and {\BBless} decay and are assumed to be uniform across the
detector. We account for the 5.3\,yr half-life of $^{60}$Co by
parameterizing $R^{\rm Co}(t) = R^{\rm Co}(0)e^{-t/\tau_{\rm Co}}$,
where $R^{\rm Co}(0)$ (a free parameter in the fit) is the $^{60}$Co
decay rate at $t=0$ which corresponds to the start of data-taking in
March 2013 and $\tau_{\rm Co}$ is the $^{60}$Co lifetime. For each
{\BoDs}, we fix the location of the {\BBless} decay signal at
$E^{0\nu}_\bd\equiv\hat\mu_\bd-87.00+\Delta\mu(Q_{\beta\beta})$, which
is the expected location of a potential {\BBless} decay signal after
correcting for the small residual calibration bias. The position of
the $^{60}$Co peak is handled identically to the other peaks in the
physics spectrum: we parameterize $E^{\rm
  Co}_\bd\equiv\hat\mu_\bd-\Delta$, with $\Delta$ left unconstrained
in the fit. We model the background as constant in energy over this
range, with $b_{\rm ROI}$ giving the rate in {\ckky}, common for all
{\BoDs} pairs and unconstrained in the fit. We test other possible
background shapes (i.e., linear and parabolic) as part of our
systematic study.

We can directly relate $R^{0\nu}$ to the physical $^{130}$Te {\BBless}
decay rate $\Gamma_{0\nu}$ through
\begin{equation}
  R^{0\nu}=\varepsilon_{0\nu\beta\beta}\frac{a_iN_A}{W}\Gamma_{0\nu},
\end{equation}
where $\varepsilon_{0\nu\beta\beta}$ is the total signal efficiency
calculated in Section~\ref{sec:CutsAndEfficiency}, $a_i$ is the
isotopic abundance of $^{130}$Te, 34.167\%, $W$ is the molar mass of
TeO$_2$ and $N_A$ is Avogadro's number.

\begin{figure}
  \centering
  \includegraphics[width=.98\columnwidth]{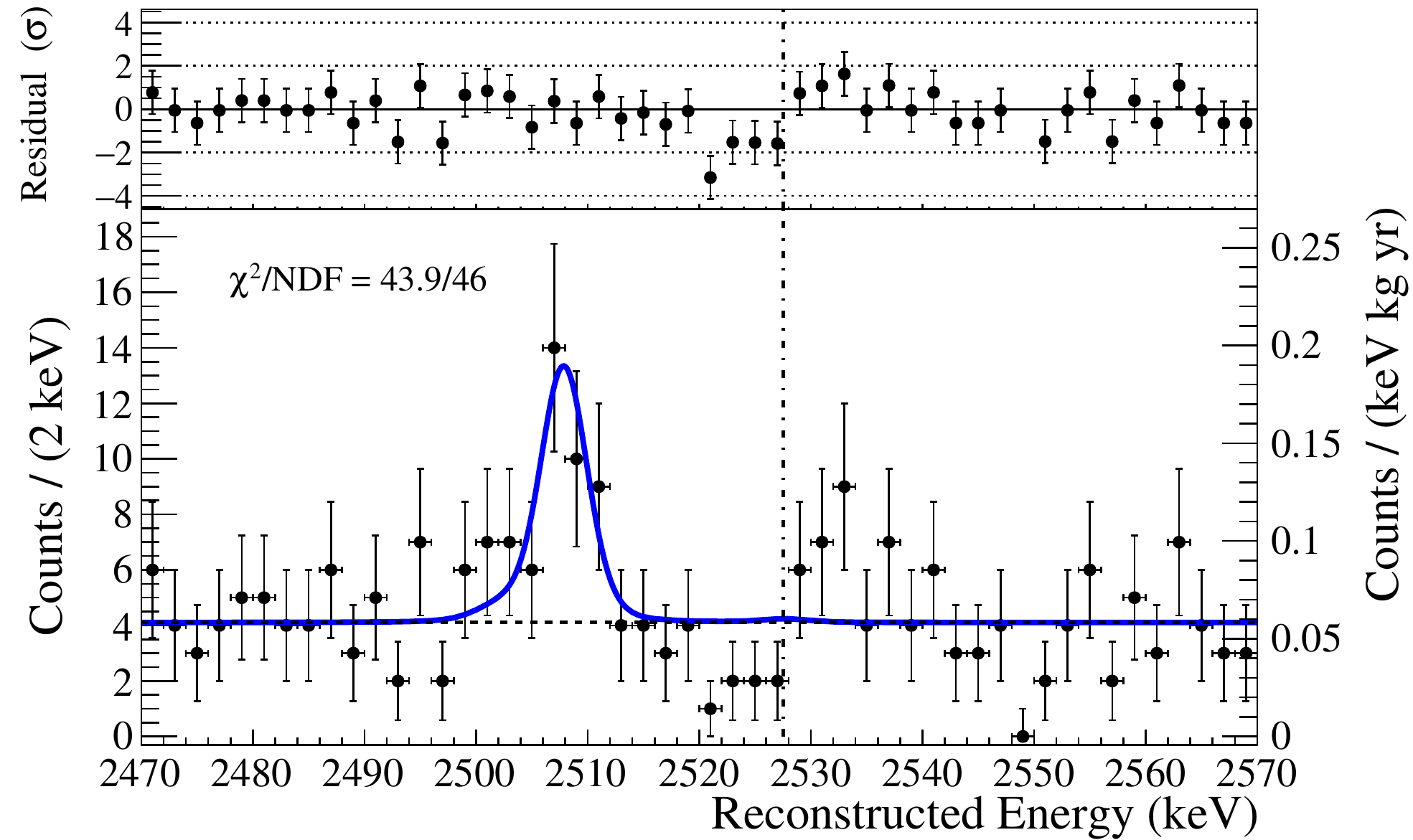}
  \caption{\emph{Bottom pannel:} The best-fit model (\emph{solid
      \grayselect{black}{blue} line}) overlaid on the {\q} energy
    spectrum (\emph{data points}). For simplicity, the data are shown
    with Gaussian error bars. The peak at 2507\,keV is due to $^{60}$Co
    and the dash-dotted line indicates the position at which we expect
    a potential {\BBless} decay signal. The dashed black line
    indicates the continuum background component in the ROI. \emph{Top
      pannel:} The normalized residuals of the best fit model and the
    binned data points. Figure from~\cite{Q0FinalPrl}.}
  \label{fig:ROIWithFit}
\end{figure}

The resulting best-fit parameters are listed in
Table~\ref{tab:BestFitROIParams} and the best-fit model is shown in
Fig.~\ref{fig:ROIWithFit}.

\begin{table}
  \centering
  \caption{The best-fit parameters from the ROI fit. The $^{60}$Co
    peak position, $E^{\rm Co}$, is constructed from the fit parameter
    $\Delta$ as $E^{\rm Co}=2614.511-\Delta$.}
  \begin{tabular}{lc}
    \hline
    \hline
    $R^{\rm Co}(0)$ & $0.92\pm0.24$\,{\cky} \\
    $E^{\rm Co}$ & $2507.6\pm0.7$\,keV \\
    $b_{\rm ROI}$ & $0.058\pm0.004$\,{\ckky}\\
    $\Gamma_{0\nu}$ &  $0.01\pm0.12\times10^{-24}$\,yr$^{-1}$\\
    \hline
    \hline
  \end{tabular}
  \label{tab:BestFitROIParams}
\end{table}

\textbf{Consistency of Model:} We perform several goodness-of-fit
tests of the model. We measure a $\chi^2$ from the binned data in
Fig.~\ref{fig:ROIWithFit} of 43.9 for 46 degrees of freedom. In a
large set of pseudo-experiments generated from the best-fit model, we
find that about 90\% of experiments return a larger $\chi^2$. We find
similar consistency according to both Kolmogorov-Smirnov and
Anderson-Darling metrics~\cite{Anderson1952}. 

We also postulate an extra signal peak at the most significant
positive fluctuation around 2535\,keV, but with the position left
unconstrained. This returns an improvement in the fit of
$\Delta\chi^2=4.72$. The probability of such a fluctuation occurring by
chance is 3\% for 1 extra degree of freedom; however the probability
of it occurring by chance anywhere in the 100\,keV ROI (i.e., the
``look-elsewhere effect'') is $\approx40\%$, so we are unable to
conclude that the fluctuation is physical. For comparison, fitting the
spectrum without a line for $^{60}$Co yields a $\Delta\chi^2=24.3$ for
2 degrees of freedom. Thus the probability of this peak occurring by
chance is $0.0005$\%.

\subsection{Systematics Accounting}
The primary sources of systematic uncertainty are listed in
Table~\ref{tab:SystematicAccounting}. We consider two types of
systematic uncertainties: a systematic scaling $\sigma_{\rm
  scaling}$ which contributes an uncertainty proportional to the true
decay rate, and an additive systematic uncertainty $\sigma_{\rm
  add}$ which is independent of the decay rate. The effect of the
uncertainty on the signal detection efficiency
$\varepsilon_{0\nu\beta\beta}$ is a straightforward scaling
uncertainty. We estimate the effect of the other uncertainties on the
measured decay rate using a large ensemble of pseudo-experiments.

For the uncertainty on the energy resolution scaling $\alpha_\sigma$
and the uncertainty on the energy scale $\Delta\mu(Q_{\beta\beta})$
we modify the parameter value by 1$\sigma$, redo the fit, and generate
a set of Monte Carlo spectra with the new best fit parameters and a
simulated {\BBless} decay rate ranging from
0--2$\times10^{-24}$\,yr. For each generated spectrum, the number of
events is Poisson distributed with the expected number of events for
that set of parameters and signal. We fit the Monte Carlo
spectra with the unmodified parameters and regress the resulting
measured decay rates against the simulated values to determine
$\sigma_{\rm add}$ and $\sigma_{\rm scaling}$ for each systematic
uncertainty.

We perform a similar procedure for the choice of background model and
detector response lineshape. For the former, we simulate spectra using
best-fit background model with either a 1st or 2nd-degree polynomial
and determine the effect on the measured decay rate. For the
lineshape, we simulate data with a single Gaussian lineshape
(i.e., with $\eta_\bd=0$).

Finally, we also take into account any potential bias introduced from
the fitting procedure itself. We calculate this bias in the same way
described above, but with no parameters modified and the number of
events fixed to 233.  The results for the considered sources of
systematic errors are summarized in
Table~\ref{tab:SystematicAccounting}.

Including systematics, our best-fit {\BBless} decay rate is
\begin{equation}
  \hat\Gamma_{0\nu}=(0.01\pm0.12\,{\rm (stat)}\pm0.01\,{\rm
    (syst)})\times10^{-24}\,\text{yr}^{-1}.
  \label{eqn:BestFit}
\end{equation}

We follow a similar procedure to calculate the systematic error on the
background rate in the ROI, $b_{\rm ROI}$, and obtain
\begin{equation}
  \begin{array}{r}
    b_{\rm ROI}=0.058\pm0.004\,{\rm (stat)}\pm0.002\,{\rm (syst)}\hspace{1cm}\;\\\text{\ckky}.
  \end{array}
  \label{eqn:BestFitBkg}
\end{equation}
Using this value, we calculate the 90\%\,C.L. sensitivity of the
experiment, as the median 90\%\,C.L. limit of a large number of MC
pseudoexperiments generated with this expected background and no
{\BBless} decay signal. The resulting 90\%\,C.L. sensitivity is
\mbox{$2.9\times10^{24}$\,yr} --- slightly surpassing the Cuoricino
sensitivity and limit~[\QINOPaper].

\begin{table}
  \centering
  \caption{Summary of the systematic uncertainties and their effect on
    the {\BBless} decay rate. Adapted from \cite{Q0FinalPrl}.}
  \begin{tabular}{lll}
    \hline
    \hline
    & Additive ($10^{-24}$\,yr$^{-1}$) & Scaling (\%)\\
    \hline
    Signal Detection & - & 0.7\\
    \hline
    Energy resolution & 0.006 & 2.6 \\
    Energy scale & 0.006 & 0.4 \\
    Bkg function & 0.004 & 0.7 \\
    Lineshape & 0.004 & 1.3 \\
    Fit bias & 0.006 & 0.15 \\
    \hline
    \hline
  \end{tabular}
  \label{tab:SystematicAccounting}
\end{table}

\subsection{Limit Evaluation}
Since our best fit value of $\Gamma_{0\nu}$ is compatible with 0, we
conclude that we see no evidence of a {\BBless} decay signal and set a
Bayesian upper limit on the {\BBless} decay rate of $^{130}$Te.  We
eliminate our nuisance parameters $\nu\equiv\left\{R^{\rm
  Co}(0),E^{\rm Co},b_{\rm ROI}\right\}$ by maximizing the likelihood
over them and calculating the likelihood ratio,
$\mathcal{L}_{\rm PR}$:
\begin{equation}
  \mathcal{L}_{\rm PR}(\Gamma_{0\nu}) \equiv \frac{\mathcal{L}|_{\rm max\,\nu}(\Gamma_{0\nu}, \nu)}{\mathcal{L}(\hat{\Gamma}_{0\nu}, \hat{\nu})},
\end{equation}
where $\hat{\Gamma}_{0\nu}$, $\hat{\nu}$ are the best-fit values. 

We evaluate our upper limit $\Gamma_{\rm Limit}$ at a confidence of
$\alpha_{\rm CL}$ as
\begin{equation}
  \alpha_{\rm CL}=\frac{\int_{-\infty}^{\Gamma_{\rm Limit}}\!\mathcal{L}_{\rm PR}(\Gamma)\pi(\Gamma)\,{\rm d}\Gamma}{\int_{-\infty}^{\infty}\!\mathcal{L}_{\rm PR}(\Gamma)\pi(\Gamma)\,{\rm d}\Gamma},
\end{equation}
where $\pi(\Gamma)$ is the prior on $\Gamma_{0\nu}$. We assume a flat
prior in the physical region, $\pi(\Gamma)=1$ for $\Gamma\geq0$ and
$\pi(\Gamma)=0$ otherwise.

We place an upper limit of
$\Gamma_{0\nu}<0.25\times10^{-24}$\,yr$^{-1}$ or
$T^{0\nu}_{1/2}>2.7\times10^{24}$\,yr at 90\%\,C.L (only accounting
for statistical uncertainties) \cite{Q0FinalPrl}. This limit is
slightly worse than our median expected 90\%\,C.L.\ sensitivity of
\mbox{$2.9\times10^{24}$\,yr} due to a slight upward fluctuation at
$Q_{\beta\beta}$. The probability to obtain a more stringent limit is
55\%.

We account for our systematic uncertainties by first combining them in
quadrature to create a single $\sigma_{\rm syst}(\Gamma)$:
\begin{equation}
  \sigma^2_{\rm syst}(\Gamma)\equiv\sum_i \sigma_{{\rm add},i}^2+\sigma_{{\rm scaling},i}^2\Gamma^2,
\end{equation}
where the sum runs over all systematic uncertainties. We include this
in our profile likelihood curve, $\mathcal{L}_{\rm PR}(\Gamma)$, using the
method outlined in \QINOPaper. We denote 
\begin{equation}
  \chi^2_{\rm stat} \equiv 2{\rm NLL}_{\rm stat} = -2\log\mathcal{L}_{\rm PR}.
\end{equation}
We assume a Gaussian distribution for our total systematic uncertainty
such that the measured $\Gamma_{0\nu}$ is normally distributed around
the best-fit value $\hat{\Gamma}_{0\nu}$, with variance $\sigma^2_{\rm
  syst}$. We construct the function \mbox{$\chi^2_{\rm
    syst}\equiv(\Gamma_{0\nu}-\hat\Gamma_{0\nu})^2/\sigma^2_{\rm
    syst}$} and combine statistical and systematic uncertainties into
a total $\chi^2_{\rm tot}$ distribution:
\begin{equation}
\frac{1}{\chi^2_{\rm tot}(\Gamma)}=\frac{1}{\chi^2_{\rm stat}(\Gamma)}+\frac{1}{\chi^2_{\rm syst}(\Gamma)}.
\end{equation}
From $\chi^2_{\rm tot}$ we can calculate the negative-log-likelihood
including systematics, ${\rm NLL}_{\rm stat+syst} \equiv
\frac{1}{2}\chi^2_{\rm tot}$. The negative log likelihoods, NLL$_{\rm
  stat}$ and NLL$_{\rm stat+syst}$, are plotted in
Fig.~\ref{fig:NLLCurves} and the difference between them is almost
negligible. At our 90\% limit, the systematic uncertainty only
accounts for 5\% of our total uncertainty, so these results are
statistics limited. Including the systematic uncertainties, we set a
limit on the {\BBless} decay rate of $^{130}$Te of
\mbox{$\Gamma_{0\nu}<0.25\times10^{-24}$\,yr$^{-1}$} or
\mbox{$T^{0\nu}_{1/2}>2.7\times10^{24}$\,yr} at 90\%\,C.L
\cite{Q0FinalPrl}.

\section{Discussion of Fit Results \& Detector Performance}
\label{sec:DetectorPerformance}
\subsection{Detector Energy Resolution}
\begin{figure}
  \centering
  \includegraphics[width=.98\columnwidth]{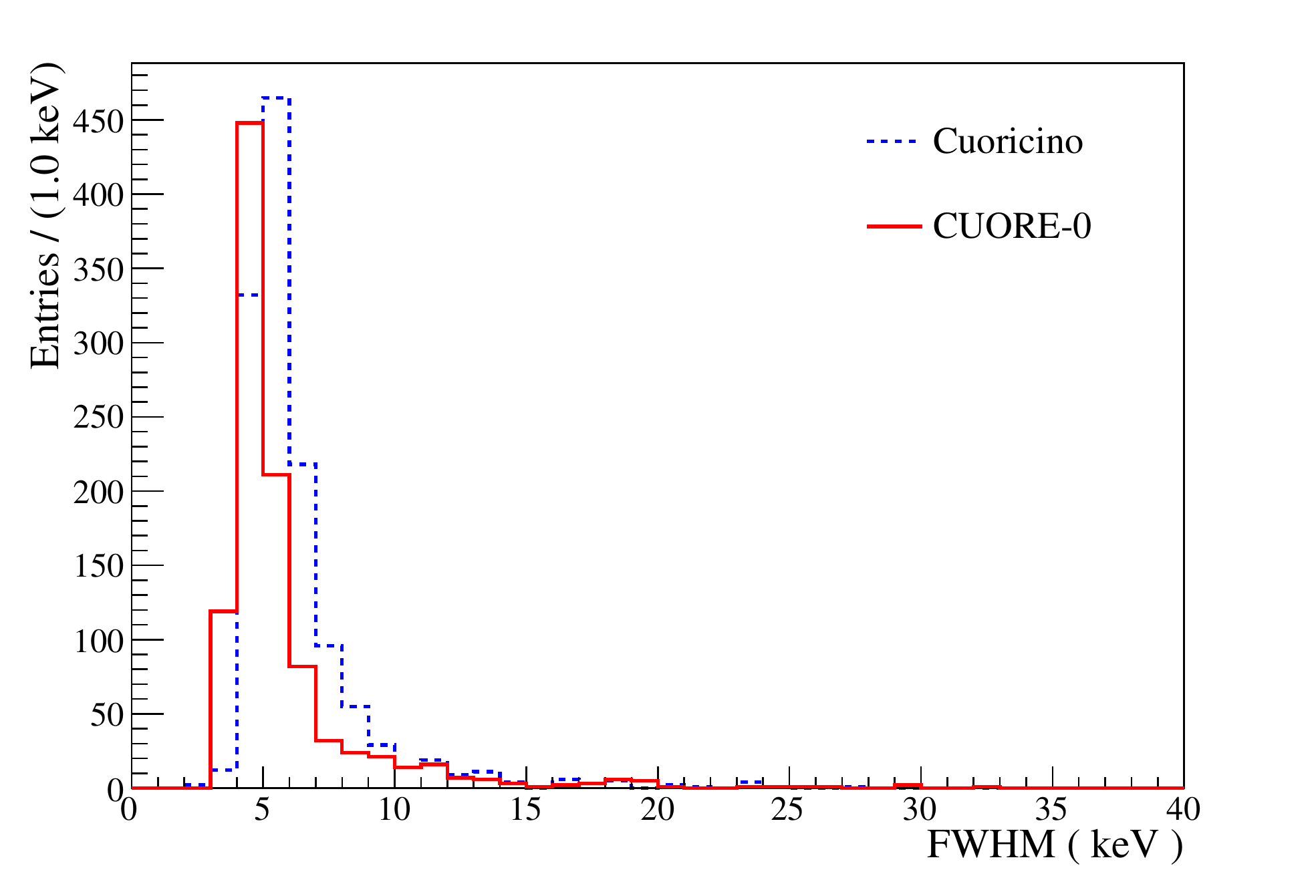}
  \caption{The distribution of FWHM values for each {\BoDs} for {\q}
    (\emph{\grayselect{gray}{red}-solid}) compared to the similar
    distribution for {\qino}
    (\emph{\grayselect{black}{blue}-dashed}).}
  \label{fig:FWHMDistribution}
\end{figure}

To characterize the energy resolution of our detector, we quote the
FWHM at the $^{208}$Tl energy in calibration runs. Each {\BoDs} has
its own best fit probability density function (PDF), for which we
numerically evaluate the FWHM. We measure the FWHM of the summed
primary and secondary peaks without the background continuum. This
yields a distribution of FWHM values, one for each {\BoDs}, which is
shown in Fig.~\ref{fig:FWHMDistribution}.

In order to quote a single FWHM value characteristic of the entire
detector performance, we calculate an effective FWHM which is obtained
by averaging detector sensitivities (i.e., we quote the resolution of a
single bolometer with equivalent sensitivity). Numerically, this is a
weighted harmonic mean of resolutions:
\begin{equation}
  \textrm{Effective FWHM} = \left.\sum_\bd T_\bd\middle/\sum_\bd \frac{T_\bd}{\Delta E_\bd}\right.,
\end{equation}
where $T_\bd$ is the physics exposure of bolometer-dataset $\bd$. For
convenience, we also quote the combined FWHM values, which are defined
as the FWHM of the combined PDF fits to the calibration data shown in
Fig.~\ref{fig:SummedChannelCalibrationFit}. The resulting
resolution values are shown in Table~\ref{tab:FWHMTable}. We also
quote the projected resolution at $Q_{\beta\beta}$ in the physics
spectrum, $5.1\pm0.3$\,keV, by multiplying the effective FWHM for the
full {\q} data by $\alpha_\sigma(Q_{\beta\beta})$.

\begin{table}
  \centering
  \caption{FWHM values for {\q} and {\qino} data measured on the
    calibration $^{208}$Tl 2615\,keV line (see text for
    details).}
  \begin{tabular}{lcc}
    \hline
    \hline
    & Combined FWHM & Effective FWHM \\
    & (keV) & (keV)\\
    \hline
    {\q} Campaign I & 5.3 & 5.7 \\
    {\q} Campaign II & 4.6 & 4.8 \\
    \hline
    {\q} Total & 4.8 & 4.9 \\
    \hline
    {\qino} & & 5.8 \\
    \hline
    \hline
  \end{tabular}
  \label{tab:FWHMTable}
\end{table}

\subsection{{\q} Background Rate}
\begin{figure}
  \centering
  \includegraphics[width=.98\columnwidth]{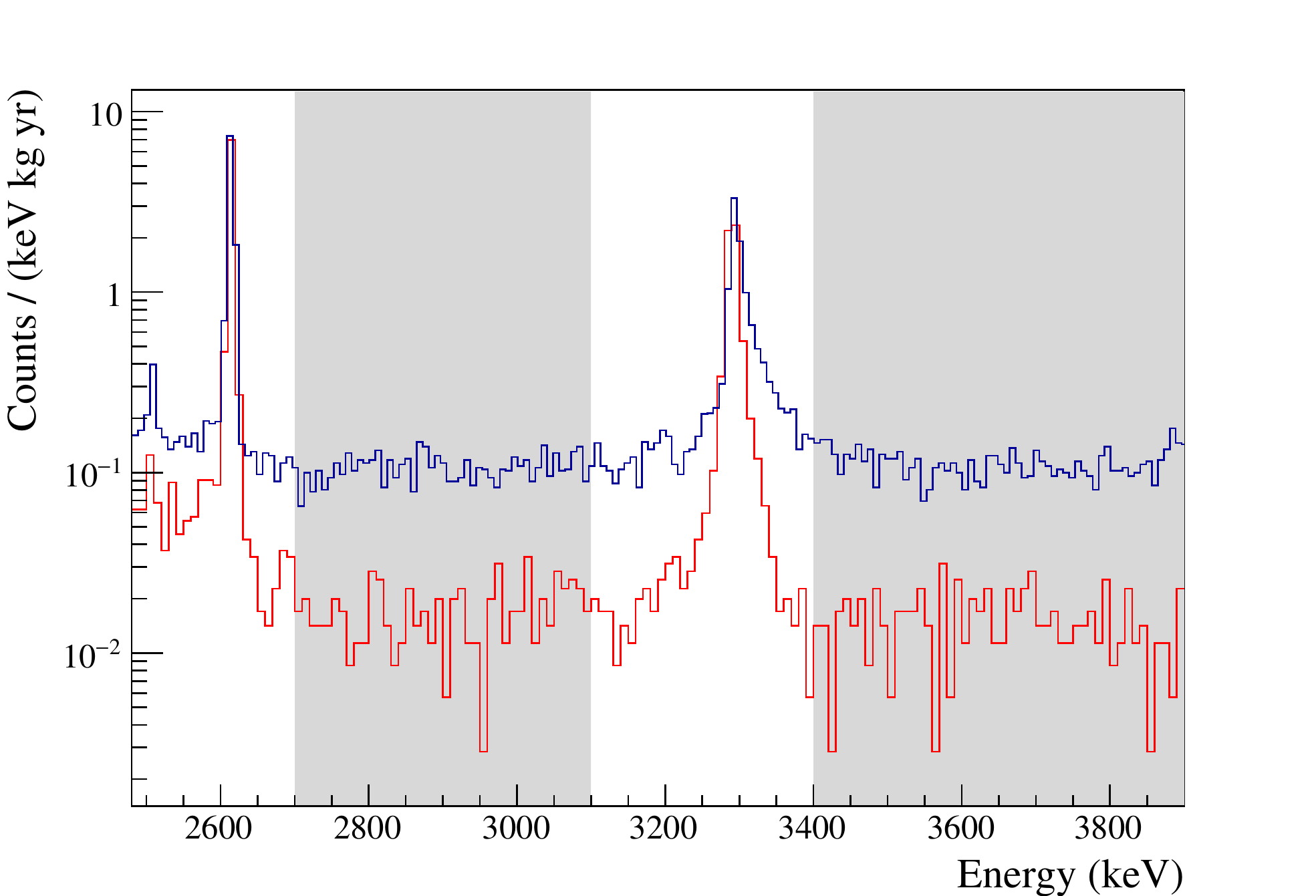}
  \caption{The $\alpha$-continuum region for {\q}
    (\emph{\grayselect{solid}{red lower} line}) and {\qino}
    (\emph{\grayselect{dotted}{blue upper} line}). The regions over
    which the $\alpha$-continuum is evaluated are the shaded regions
    from 2700--3900 excluding the $^{190}$Pt peak. Note that the
    $^{190}$Pt peak reconstructs $\sim40\,\rm{keV}$ too high due to
    the quenching of $\alpha$-particles in the bolometers; this is
    discussed further in \cite{Q0BackgroundModel}.}
  \label{fig:AlphaContRegion}
\end{figure}

In order to quantify the background reduction achieved relative to
{\qino}, and to compare with the projections for CUORE, we consider
the background rate in two regions of the spectrum: the
$\alpha$-region and the ROI.  The $\alpha$-background is measured over
the range 2700--3900\,keV, which is dominated by degraded alpha
decays. We exclude the range 3100--3400\,keV which contains a peak
from the decay of $^{190}$Pt (see
Fig.~\ref{fig:AlphaContRegion}). $^{190}$Pt is a naturally occurring
isotope that contaminates our bolometers during the crystal growth
process. However, since the contamination is usually in the form of
inclusions within the crystal bulk, the $\alpha$-particles emitted in
these decays do not degrade and thus do not contribute to the
background in the ROI. We also see no evidence for a $^{190}$Pt peak
in the two-crystal coincidence spectrum, which indicates that the
contamination is not near the surface~\cite{Q0BackgroundModel}. Over
the $\alpha$-continuum range, we measure an average rate of
\mbox{$b_{\alpha}=0.016\pm0.001$\,{\ckky}}. This is in agreement with
our projected background for {\q}. Comparing this to the value
measured in {\qino}, \mbox{$b_{\alpha}=0.110\pm0.001$\,{\ckky}}, we
see an improvement of a factor of 6.8.

The background in the ROI is expected to have both an $\alpha$
component which extends down from the $\alpha$ region discussed above
and a $\gamma$ component from $^{208}$Tl 2615\,keV $\gamma$-rays which
undergo low angle scattering before being absorbed in a bolometer or
multiple scattering in a single bolometer before escaping. We measure
the background rate in the ROI from the UEML fit to the ROI,
\mbox{$b_{\rm ROI}=0.058\pm0.004$\,{\ckky}}. This is an improvement of
a factor of 2.7 over {\qino}, which obtained \mbox{$b_{\rm
    ROI}=0.153\pm0.006$\,{\ckky}} in its similarly sized unenriched
crystals [\QINOPaper]. Taking the difference between $b_{\rm ROI}$ and
$b_{\alpha}$, we can estimate the $\gamma$ background component to be
\mbox{$b_{\gamma}=0.042\pm0.004$\,{\ckky}} and
\mbox{$b_{\gamma}=0.043\pm0.006$\,{\ckky}} for {\q} and {\qino}
respectively. This is consistent with our models that place the origin
of the $\gamma$ contamination in the cryostat materials, which are
common to both {\q} and {\qino}~\cite{Q0BackgroundModel}.  This
$\gamma$ background forms an irreducible background for {\q}, but is
expected to be significantly reduced in {\Q} due to better material
selection, better shielding, and more efficient anti-coincidence
rejection.

Overall, {\q} successfully validated the background reduction measures
developed for {\Q} and we believe that the {\Q} background goal of
\mbox{$b_{\rm ROI}=0.01$\,{\ckky}} is within reach. A projection of the
     {\q} results to the {\Q} background will be detailed in a paper
     currently in preparation~\cite{Q0BackgroundModel}.

\subsection{Position of the $^{60}$Co Sum Peak}
\begin{figure}
  \centering
  \includegraphics[width=.98\columnwidth]{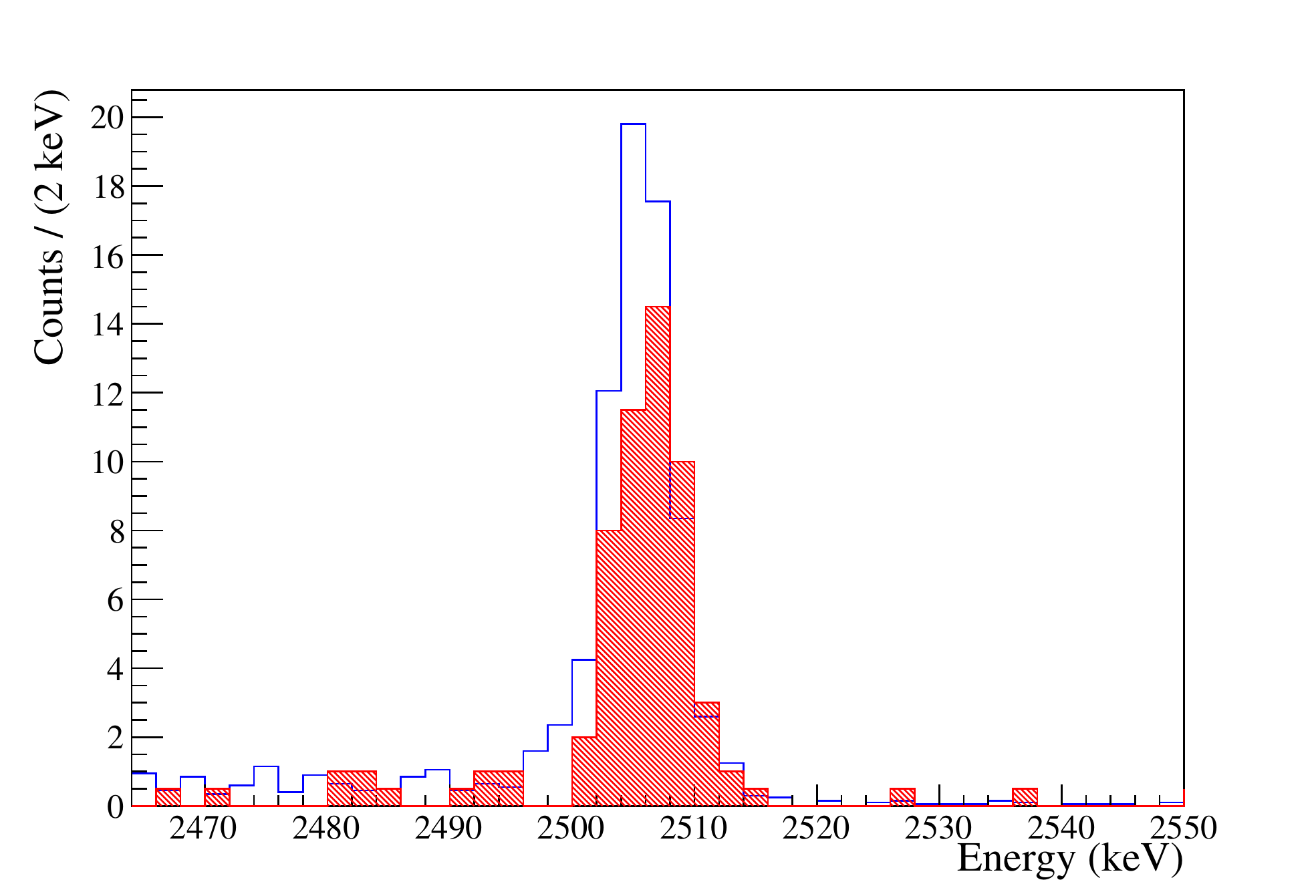}
  \caption{The spectrum from the dedicated $^{60}$Co calibration
    measurement. The single-crystal coincidence peak
    (\emph{\grayselect{gray}{red} shaded}) reconstructs 2\,keV higher
    than the summed energy peak from the two-crystal coincidence
    (\emph{\grayselect{black}{blue}}).}
  \label{fig:Co60Plot}
\end{figure}

In Table~\ref{tab:BestFitROIParams}, we observe that the
single-crystal $^{60}$Co coincidence peak reconstructs
$1.9\pm0.7$\,keV higher than expected. A similar effect was seen in
{\qino}, where the peak reconstructed too high by $0.8\pm0.3$\,keV
(using a different lineshape and fitting procedure). In {\QINOPaper},
we took this shift as a systematic uncertainty on our {\BBless} peak
position; however, in Fig.~\ref{fig:RelativeResiduals} this peak is a
clear outlier.  We confirmed this effect with a dedicated calibration
measurement with $^{60}$Co sources. The total energy from a
two-crystal coincidence --- when the two $\gamma$-rays are absorbed in
two distinct crystals --- reconstructs at the expected energy whereas
the single-crystal coincidence reconstructs too high by 2\,keV, see
Fig.~\ref{fig:Co60Plot}.

At present, we are still investigating the cause of this effect, but
we hypothesize that it is due to a difference in the geometric spread
in the energy deposition in the crystal. A coincidence of two
$\gamma$-rays in a single crystal can spread the energy over a larger
volume of the bolometer, and this may affect the bolometric gain. The
difference in energy deposition topologies is borne out in MC
simulations, however the mechanism connecting this to the gain is
still uncertain. This hypothesis is supported by the fact that the
single-escape peak from $^{208}$Tl at 2104\,keV also reconstructs
significantly higher than expected, by $0.84\pm0.22$\,keV. This peak
is caused by pair production in a crystal, followed by an annihilation
to two 511\,keV $\gamma$-rays, one of which is reabsorbed in the same
crystal. \edit{The double escape peak in the physics data at 1593\,keV
  reconstructs within $0.13\pm0.30\,\mathrm{keV}$ of the expected
  value, following the trend of the other peaks in the physics
  spectrum. We conclude from this that these events, and consequently
  the {\BBless} decay events which are topologically similar, are not
  shifted by the same effect as the $^{60}$Co single crystal
  coincidence. We note however, that in the calibration spectrum, this
  peak reconstructs low by $-0.66\pm0.02\,\mathrm{keV}$. We continue
  to investigate the origin of this discrepancy.}

\section{Combination with {\qino}}
\label{sec:CombinedLimits}
In this section we combine the results of {\q} with those of {\qino}
(19.75\,{\kgyr} of $^{130}$Te). The {\qino}
experimental design and data analysis has already been addressed in
this document and the details and final {\BBless} decay results can be
found in {\QINOPaper}.

\begin{figure}
  \centering
  \includegraphics[width=.98\columnwidth]{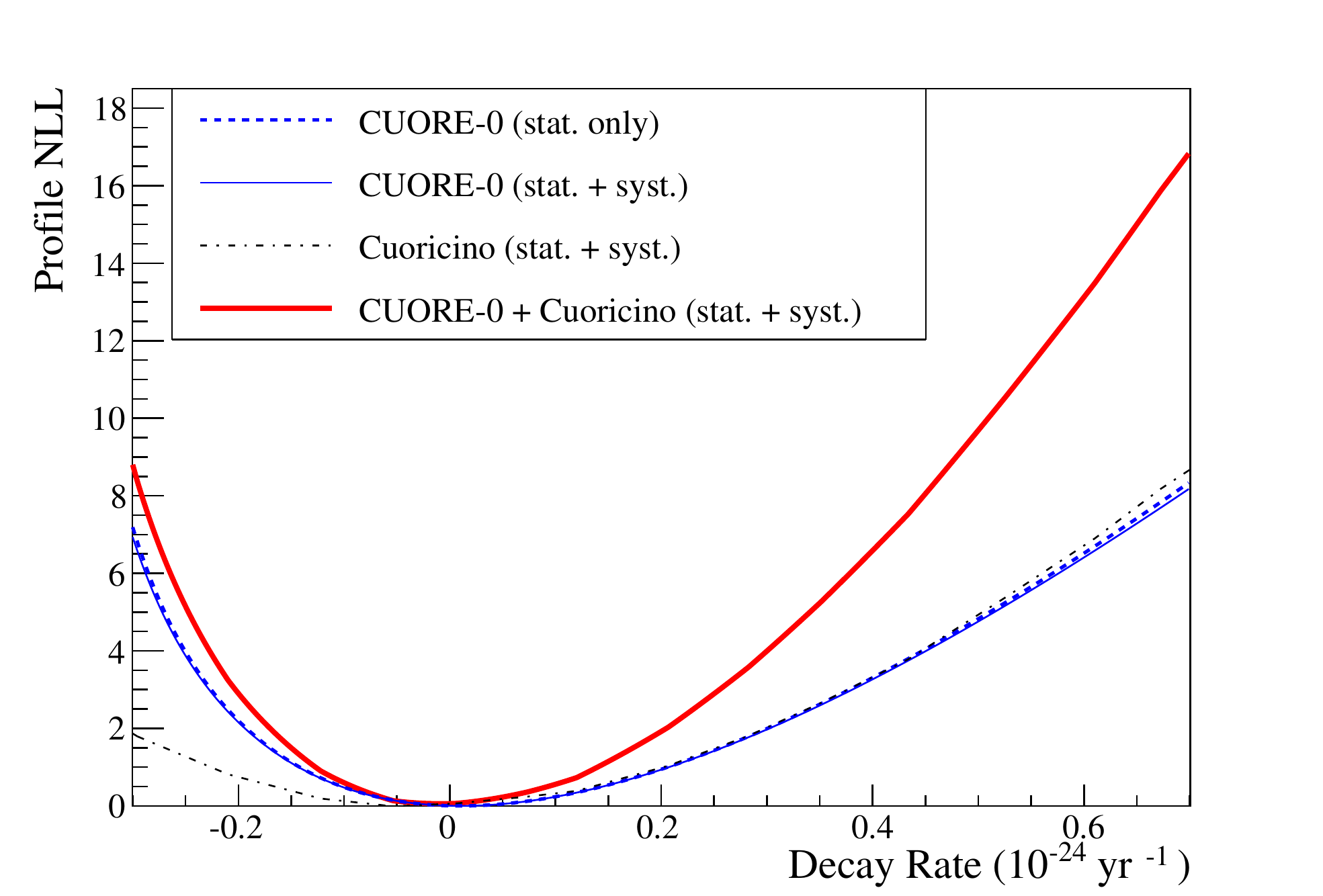}
  \caption{Combined negative log-likelihood (NLL) curves for {\q},
    {\qino}, and their combined curves. Figure
    from~\cite{Q0FinalPrl}.}
  \label{fig:NLLCurves}
\end{figure}

We combine the {\q} likelihood curve with the one from
{\qino}\footnote{We point out that for the purposes of combining the
  results, the {\qino} profile likelihood curve has been updated to
  reflect a new measurement of the $^{130}$Te isotopic abundance,
  which changed from $a_I=0.3380(1)$ to
  $a_I=0.341668\pm0.000016$~\cite{Fehr2004}; however, all {\qino}
  results quoted here continue to reflect the old measurement.} (see
Fig.~\ref{fig:NLLCurves}) and set a combined limit on the {\BBless}
decay rate of $^{130}$Te at
\mbox{$\Gamma_{0\nu}<0.17\times10^{-24}$\,yr$^{-1}$} or
\mbox{$T_{1/2}^{0\nu}>4.0\times10^{24}$\,yr} at 90\%\,C.L
\cite{Q0FinalPrl}. 

In addition to our Bayesian limit, we also report a frequentist limit
using the Rolke technique~\cite{Rolke2005}. For {\q} we obtain
\mbox{$T^{0\nu}_{1/2}>2.8\times10^{24}$\,yr} including systematic
uncertainties and \mbox{$T^{0\nu}_{1/2}>4.1\times10^{24}$\,yr} for the
combined bolometric limit.

\begin{figure}[h!]
  \centering
  \includegraphics[width=.98\columnwidth]{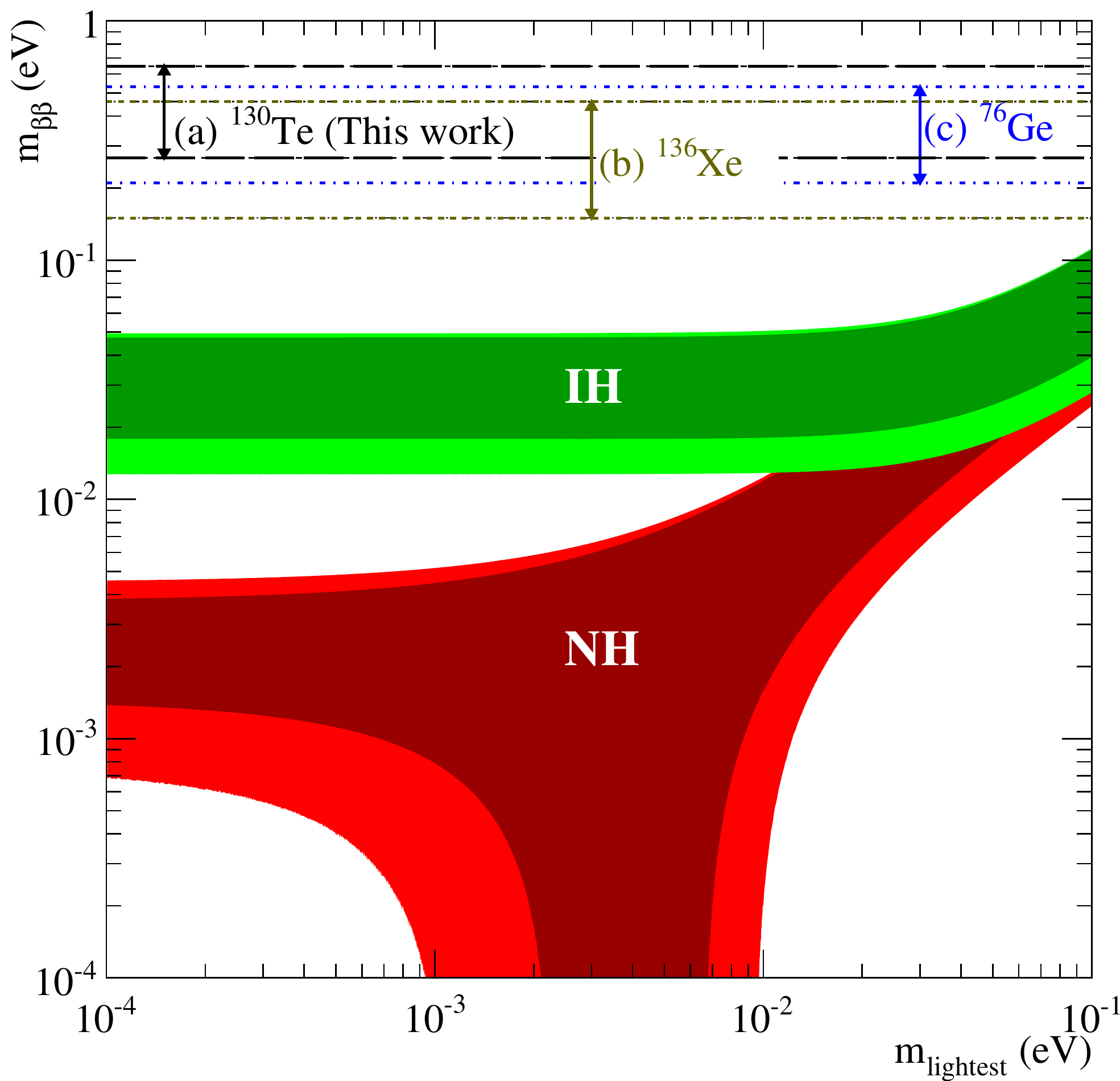}
  \caption{Plot of the allowed regions for $m_{\beta\beta}$ as a
    function of the lightest neutrino mass, $m_{\rm lightest}$,
    assuming the light Majorana neutrino exchange model of {\BBless}
    decay. The exclusion bands represent 90\%\,C.L., with the vertical
    width coming from the uncertainty in the NME. The exclusion from
    $^{76}$Ge is from~\cite{GERDA2013}. The exclusion from $^{136}$Xe
    is the combined limits from~\cite{KamLANDZen2013,EXO2014}. The
    exclusion from $^{130}$Te is the result of this paper. Figure
    from~\cite{Q0FinalPrl}.}
  \label{fig:LobsterPlot}
\end{figure}

We interpret our combined Bayesian half-life limit in the context of
{\BBless} decay mediated by the exchange of light Majorana
neutrinos~\cite{Kotila2012} and place a limit on the effective
Majorana mass, $m_{\beta\beta}$, where $m_{\beta\beta}$ is defined by
\begin{equation}
  \frac{1}{T^{0\nu}_{1/2}}=G_{0\nu}\left|\mathcal{M}_{0\nu}\right|^2\frac{m_{\beta\beta}^2}{m_e^2}.
\end{equation}
Using the phase space factors $G_{0\nu}$ from~\cite{Kotila2012}, and
assuming a value of $g_A\simeq1.27$, we can place the following range
of limits for the most recent nuclear matrix element (NME),
$\left|\mathcal{M}_{0\nu}\right|^2$, calculations:
\begin{itemize}
\item $m_{\beta\beta}<520-650\,\rm{meV}$ for Interacting Shell Model
  calculations from~\cite{Menendez2009};
\item $m_{\beta\beta}<300-340\,\rm{meV}$ for the Quasiparticle-Random
  Phase Approximation (QRPA) calculation from~\cite{Simkovic2013};
\item $m_{\beta\beta}<340\,\rm{meV}$ for the QRPA calculation
  from~\cite{Hyvrinen2015};
\item $m_{\beta\beta}<360\,\rm{meV}$ for the Interacting Boson Model
  calculation from~\cite{Barea2015};
\item $m_{\beta\beta}<270\,\rm{meV}$ for the Energy Density Functional
  calculations from~\cite{Rodriguez2010};
\item $m_{\beta\beta}<700-760\,\rm{meV}$ for the shell model
  from~\cite{Neacsu2015}.
\end{itemize}
Using all available calculations, we place a limit range on the
effective Majorana mass of $m_{\beta\beta}<270-760\,\rm{meV}$
\cite{Q0FinalPrl}. However, since the shell model calculations
from~\cite{Neacsu2015} are not presently available for other isotopes,
we present an alternative limit useful for comparison with limits from
$^{76}$Ge and $^{136}$Xe,
\mbox{$m_{\beta\beta}<270-650\,\rm{meV}$}. This latter limit is
presented in Fig.~\ref{fig:LobsterPlot}.

\section{Conclusions}

The {\q} experiment ran for two years from 2013 to 2015 and collected
35.2\,{\kgyr} of TeO$_2$ exposure or 9.8\,{\kgyr} of $^{130}$Te
exposure. The improved background level of the detector allows us to
reproduce the sensitivity of the {\qino} experiment in $\sim$40\% the
time. In this paper, we fully describe the procedure adopted for the
analysis of this data that was presented in \cite{Q0FinalPrl}, which
placed a 90\%\,C.L.\ Bayesian limit on the {\BBless} decay half-life
of $^{130}$Te at \mbox{$T^{0\nu}_{1/2}>2.7\times10^{24}$\,yr} and a
combined limit with the {\qino} data of
\mbox{$T^{0\nu}_{1/2}>4.0\times10^{24}$\,yr}. This corresponds to a
limit range on the effective Majorana mass of
\mbox{$m_{\beta\beta}<270-760\,\rm{meV}$}, using the most up-to-date
NME calculations.

These results also validate the techniques we have developed in
preparation for {\Q}. We have achieved the {\Q} energy resolution goal
of 5\,keV FWHM at 2615\,keV. We also achieved a background of
\mbox{$b_{\rm ROI}=0.058\pm0.004$\,{\ckky}} in the ROI and
\mbox{$b_{\alpha}=0.016\pm0.001$\,{\ckky}} in the $\alpha$-continuum,
which are in line with our predictions and gives us confidence that the
{\Q} goal is within reach. We have developed an algorithm to improve
the energy resolution by deconvolving signals from multiple
bolometers, and a TGS algorithm to recover data from bolometers with
missing Joule heaters. We have improved the efficiency of our
anti-coincidence cut, which will be necessary in the larger {\Q}
detector. Finally, we have implemented a data blinding technique that
is both robust and effective.

This analysis has also highlighted several open issues that will be
addressed for {\Q}. As a result of our improved energy resolution, we
saw a more complicated lineshape than previously seen. We seek to
understand the source of this substructure and its effect on our
expected {\BBless} decay signal. We found a small energy-dependent
bias in our energy reconstruction at the level of 0.1\,keV that we
seek to address in {\Q}. Finally, we have seen a significant shift in
the reconstructed energy of the single-crystal coincidence peaks,
$^{60}$Co and the $^{208}$Tl single escape. Moving forward, we plan to
investigate the sources of these effects and their impacts on our
{\BBless} analysis as we push to better understand our detectors.
 
\section*{Acknowledgements}
The CUORE Collaboration thanks the directors and staff of the
Laboratori Nazionali del Gran Sasso and the technical staff of our
laboratories. The authors wish to thank J.~Feintzeig for carefully
reviewing the manuscript, M.~Nastasi for preparing the $^{60}$Co
calibration sources, and F.~Iachello for helpful discussions
concerning the NME literature. This work was supported by the Istituto
Nazionale di Fisica Nucleare (INFN); the National Science Foundation
under Grant Nos. NSF-PHY-0605119, NSF-PHY-0500337, NSF-PHY-0855314,
NSF-PHY-0902171, NSF-PHY-0969852, NSF-PHY-1307204, NSF-PHY-1314881,
NSF-PHY-1401832, and NSF-PHY-1404205; the Alfred P. Sloan Foundation;
the University of Wisconsin Foundation; and Yale University. This
material is also based upon work supported by the US Department of
Energy (DOE) Office of Science under Contract Nos. DE-AC02-05CH11231,
DE-AC52-07NA27344, and DE-SC0012654; and by the DOE Office of Science,
Office of Nuclear Physics under Contract Nos. DE-FG02-08ER41551 and
DE-FG03-00ER41138.  This research used resources of the National
Energy Research Scientific Computing Center (NERSC).

\end{document}